\documentclass[journal]{IEEEtran}
\usepackage{cite}
\usepackage{amsmath,amssymb,amsfonts,amsthm}
\usepackage{mathtools}
\usepackage{algorithmic}
\usepackage{graphicx}
\usepackage{textcomp}
\usepackage[T1]{fontenc}
\usepackage{caption}
\usepackage{subcaption}
\usepackage{epstopdf}
\usepackage{overpic}
\usepackage{array}
\usepackage[monochrome]{color}
\usepackage{url}
\usepackage{bm}
\usepackage{bbm}

\usepackage{hyperref}
\usepackage{lipsum}
\usepackage{overpic}

\def\BibTeX{{\rm B\kern-.05em{\sc i\kern-.025em b}\kern-.08em
    T\kern-.1667em\lower.7ex\hbox{E}\kern-.125emX}}
    
\usepackage{color}

\newcommand{\Ttran}{\rm \scriptscriptstyle T}
\newcommand{\Htran}{\rm \scriptscriptstyle H}
\def\CN{\mathcal{N}_{\mathbb{C}}}

\def\Real{\mathbb{R}}
\def\Complex{\mathbb{C}}
\def\Integer{\mathbb{Z}}

\def\Ex{\mathbb{E}}
\def\sinc{\mathrm{sinc}}
\def\rank{\mathrm{rank}}

\def\diag{\mathrm{diag}}

\def\imagunit{{\sf j}} 

\newcommand{\vect}[1]{{\boldsymbol{#1}}}

\theoremstyle{plain}

\graphicspath{{./images/}}

\begin{document}

\title{Mutual Coupling in Holographic MIMO: Physical Modeling and Information-Theoretic Analysis}

\definecolor{alizarin}{rgb}{0.82, 0.1, 0.26}
\definecolor{ao}{rgb}{0.0, 0.5, 0.0}
\newcommand{\angel}[1]{\noindent { {{$\blacktriangleright$ 
   {\textsf{[Angel]: {\color{red}#1}}} $\blacktriangleleft$}}}}
\newcommand{\andrea}[1]{\noindent { {{$\blacktriangleright$ 
   {\textsf{[Andrea]: {\color{ao}#1}}} $\blacktriangleleft$}}}}
   
\author{
\IEEEauthorblockN{Andrea Pizzo, \emph{Member, IEEE} and Angel Lozano, \emph{Fellow, IEEE}}
\thanks{
A.~Pizzo and A.~Lozano are with the Department of Engineering, Universitat Pompeu Fabra, 08018 Barcelona, Spain (andrea.pizzo@upf.edu, angel.lozano@upf.edu).
Work supported by MICIU under the Maria de Maeztu Units of Excellence Programme (CEX2021-001195-M), by ICREA, and by AGAUR (Catalan Government).
}
}


\maketitle

\begin{abstract}
This paper presents a comprehensive framework for holographic multiantenna communication, a paradigm that integrates both wide apertures and
 closely spaced antennas relative to the wavelength. The presented framework is physically grounded, enabling information-theoretic analyses that inherently incorporate correlation and mutual coupling among the antennas. 
This establishes the combined effects of correlation and coupling on the information-theoretic performance limits across SNR levels.
Additionally, it reveals that, by suitably selecting the individual antenna patterns, mutual coupling can be harnessed to either reinforce or counter spatial correlations as appropriate for specific SNRs, thereby improving the performance.
\end{abstract}

\smallskip

\IEEEpeerreviewmaketitle

\section{Introduction}


The progress of multiple-input multiple-output (MIMO) communication entails expanding the array apertures relative to the wavelength, 
so as to 
augment the number of spatial dimensions 
\cite{10144733}.
This is accomplished by physically enlarging the apertures and/or by increasing the carrier frequency, with antenna spacings at or above a half-wavelength as dictated by the channel's scattering richness \cite{BJORNSON20193} and in line with the sampling theorem \cite{PizzoTSP21}.
However, both approaches face challenges and are ultimately curbed: exceedingly large apertures become difficult to implement and deploy, while escalating frequencies compromise the range.
Further progress may be possible by shrinking the antenna spacing below 
the Nyquist distance, an idea that defines the holographic MIMO
paradigm \cite{Prather2016}. 

Antennas are inherently exposed to their mutual radiations, with the intensity of the ensuing mutual coupling being determined by their radiation patterns and relative spacing \cite{JensenReview}. 
In particular, the coupling surges as the spacing falls 
below the Nyquist distance, which is precisely the defining attribute of holographic MIMO. 
Therefore, while uncoupled formulations have been valuable in understanding fading correlation among the antennas \cite{PizzoIT21,PizzoJSAC20,PizzoTWC22}, 
ignoring coupling is ill-advised in holographic MIMO.
In recognition of this reality,
this paper introduces a unifying linear system-theoretic framework that 
intrinsically incorporates both correlation and coupling among the transmitting antennas. 
Coupling among receiving antennas would enter the formulation similarly, as it stems from the same physical phenomenon.
A fundamental analogy is uncovered between fading correlation and coupling, both manifesting as linear filtering operations, but in stark contrast. To wit, i) coupling and fading behave oppositely, with fading mapping directly onto the channel response through a spatial convolution while coupling entails a deconvolution, and ii) the wavenumber response is dictated by scattering in the case of fading, while it is tied to the antenna patterns in the case of coupling.
The two phenomena, correlation and coupling, are subsumed in a cascaded filter response. 
The present study shows how coupling can be harnessed, by suitably altering the antenna pattern, to improve this cascaded filter response and thus the performance at every signal-to-noise-ratio (SNR).
Specifically matching/countering the antenna pattern to the fading spectrum whitens/colors the cascaded wavenumber response, leading to spatial decorrelation/correlation. 
A reduced correlation is tantamount to a larger antenna spacing, effectively opening up additional spatial dimensions for multiplexing or diversity.
Antenna density is then dictated by the sampling theorem for the enlarged aperture, resulting in faster-than-Nyquist spatial sampling for the actual aperture.
The potential for such reconfigurability is backed by the latest advancements in antenna technology, including tightly coupled antennas \cite{Prather2013,Prather2017} and metasurface antennas \cite{Insang2019}.
The specific contributions of this work are as follows.
\begin{itemize}
\item
Establish a formal relationship between coupling and correlation, 
subsuming and generalizing previous conceptions of mutual coupling as a mechanism for antenna decorrelation  \cite{Steyskal1990,Birtcher2006}. 
\item
Augment the Fourier analysis in \cite{PizzoIT21,PizzoJSAC20} to include mutual coupling in wide-sense stationary spatial random channels using an accessible framework based on linear system theory and Fourier transforms.
\item
Derive the capacity-achieving transmit precoder at any arbitrary SNR, extending superdirective designs for low SNRs \cite{Wallace2005,Marzetta2019,Matthaiou2023}.
\item
Analyze how coupling affects power and spatial dimensionality, which dominate the channel capacity in the low- and high-SNR regimes, respectively.
\item
Provide additional evidence for the potential benefits of antenna densification within a fixed aperture, associated with a constructive exploitation of mutual coupling \cite{Wallace2004,Clerckx2007,Nossek2010,Masouros2013,Heath2023}, which could otherwise degrade performance  \cite{Janaswamy2002}.
\end{itemize}
Research in holographic MIMO often relies on multiport network models, where each antenna serves as a port for transmitting or receiving signals \cite{Branislav2024,Sanguinetti2024,Sha2023,Tengjiao2022,Tengjiao2023,Yongxi2024}. These models use a response matrix to capture port interactions through scattering \cite{Wallace2004} or impedance \cite{Nossek2010} formulations, relating radio signals or terminal voltages and currents among antennas.
As an alternative, one can leverage that, as antenna spacings shrink, currents on the array form a quasi-continuous holographic profile, with interactions described by a spatial impulse response.
That invites the Fourier-based framework adopted by this paper.
Both approaches rely on impedance matching networks that mix information signals at each link end, introducing another layer of interaction. Their effects are deliberately excluded here to extract insights, leaving room for further research.

The manuscript is organized as follows.
Sec.~\ref{sec:multiport} introduces the circuit model. 
Sec.~\ref{sec:coupling_iso} characterizes coupling with punctiform antennas, extended to arbitrary antennas in Sec.~\ref{sec:coupling}.
Sec.~\ref{sec:MIMO_coupled} derives the channel model with transmit coupling. 
Sec.~\ref{sec:holo_MIMO_uncoupled} revisits the Fourier channel model for uncoupled antennas, with coupling incorporated in Sec.~\ref{sec:holo_MIMO_coupled}.
Sec.~\ref{sec:DOF_increase} evaluates the impact of coupling on DOF, with 
information-theoretic performance limits studied in Sec.~\ref{sec:spectral_efficiency}.
Sec.~\ref{sec:conclusion} concludes with discussions and potential extensions.
   
\emph{Notation:} 
Lower (upper) case letters denote spatial (wavenumber) entities and calligraphic letters indicate linear operators. 
The Hilbert-Schmidt space of square-integrable functions over $\Real^3$ is denoted by $L^2$, with inner product $\langle f,g \rangle = \iiint_{-\infty}^\infty f(\vect{{\sf r}}) \overline{g(\vect{{\sf r}})} \, d\vect{{\sf r}}$ and norm $\|g\| = \sqrt{\langle g,g \rangle}$. 
Here, the adjoint $\mathcal{H}^*: L^2 \to L^2$ of an operator $\mathcal{H}: L^2 \to L^2$ is defined by the requirement $\langle f,\mathcal{H} g \rangle = \langle \mathcal{H}^* g,f \rangle$ for every $f,g$. An operator is self-adjoint when $\mathcal{H}^* = \mathcal{H}$.
In turn, 
$\mathbbm{1}_{X}$ is the indicator function of set ${X} $, $|{X}|$ is the Lebesgue measure, $\nabla$ is the gradient operator, $\nabla^2$ is the laplacian operator, ${\rm supp}(f)$ is the support of $f(\vect{\cdot})$, and $(x)^+ = \max(x,0)$. 
Also, $\vect{x}\sim\CN(\vect{\mu},\vect{Q})$ is shorthand for circularly symmetric complex Gaussian random vectors with mean $\vect{\mu}$ and covariance $\vect{Q}$.

\section{Multiantenna Circuit Model} \label{sec:multiport}

\begin{figure}
\centering\vspace{-0.0cm}
\includegraphics[width=.8\linewidth]{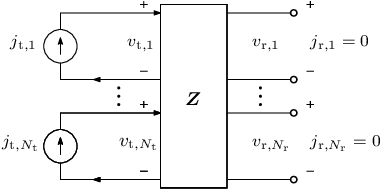} 
\caption{Multiantenna circuit model. Each transmit antenna is driven by an ideal zero-impedance current source, which is the Norton equivalent of a source with an infinite impedance in parallel. In turn, each receive antenna is connected to an infinite-impedance voltmeter, such that no current is drawn from antennas.}
\label{fig:multiport}
\end{figure}

\subsection{Mutual Impedance 
}

For narrowband communication at frequency $\omega$, each antenna is identified by a pair
of complex phasors $(j_n, v_n)$, i.e., a port, characterizing the instantaneous current and voltage via
\begin{align} \label{phasor_port}
j_n(t) = \Re\{j_n e^{-\imagunit \omega t}\} \qquad 
v_n(t) = \Re\{v_n e^{-\imagunit \omega t}\}.
\end{align}
Circuit theory inherently accounts for the interaction among antennas. Precisely, when a current is injected at one port, voltages appear across all ports.
This effect is measured by the mutual impedance \cite{Wallace2004}
\begin{equation} \label{mutual_coupling_coeff}
z_{nm} = \frac{v_n}{j_m}
\end{equation}
defined as the ratio of the open-circuit voltage at a port $m$ to the ideal current inserted at another port $n$, when all other ports are open-circuited (i.e., connected to an infinite load). The self-impedances correspond to $m=n$.
\subsection{Discrete Multiport System Model}  \label{sec:multiport_model}


A system of $N_\text{t}$ transmit and $N_\text{r}$ receive
antennas can be modeled as an $(N_\text{t}+N_\text{r})$-port network, 
specified by an impedance matrix $\vect{Z}$ 
of entries
$z_{nm}$ \cite{Nossek2010,Wallace2004}.
Transmit antenna ports are indexed by $1$ to $N_\text{t}$ and receive antenna ports by $N_\text{t}+1$ to $N_\text{t}+N_\text{r}$. Defining $\vect{v}_{\text{t}} = (v_{\text{t},1}, \ldots, v_{\text{t},N_\text{t}})^{\Ttran}$ and $\vect{j}_{\text{t}} = (j_{\text{t},1}, \ldots, j_{\text{t},N_\text{t}})^{\Ttran}$ as the transmit voltage and current vectors, and $\vect{v}_{\text{r}} = (v_{\text{r},1}, \ldots, v_{\text{r},N_\text{r}})^{\Ttran}$ and $\vect{j}_{\text{r}} = (j_{\text{r},1}, \ldots, j_{\text{r},N_\text{r}})^{\Ttran}$ as their receive counterparts, 
the multiport model partitions as \cite{Nossek2010,Wallace2004} 
\begin{equation} \label{input_output}
\begin{pmatrix}
\vect{v}_{\text t} \\
\vect{v}_{\text r}
\end{pmatrix}
=
\begin{pmatrix}
\vect{Z}_{\text{t,t}} &  \vect{Z}_{\text{t,r}} \\
\vect{H} &  \vect{Z}_{\text{r,r}}
\end{pmatrix}
\begin{pmatrix}
\vect{j}_{\text t} \\
\vect{j}_{\text r}
\end{pmatrix}
\end{equation}
where $\vect{Z}_{\text{t,t}}$ and $\vect{Z}_{\text{r,r}}$ are the transmit and receive impedance matrices, respectively, and $\vect{Z}_{\text{t,r}}$ and $\vect{H}$ the transimpedance matrices.
The latter depend on both the array geometries and the propagating medium. Due to reciprocity, a real symmetry occurs, namely $\vect{Z}_{\text{t,t}} = \vect{Z}_{\text{t,t}}^{\Ttran}$, $\vect{Z}_{\text{r,r}} = \vect{Z}_{\text{r,r}}^{\Ttran}$, and $\vect{Z}_{\text{t,r}} = \vect{H}^{\Ttran}$.
As illustrated in Fig.~\ref{fig:multiport},
in its simplest form, \eqref{input_output} reduces to \cite{MarzettaIT}
\begin{align} \label{transmit_unilateral}
\vect{v}_{\text t} & =  \vect{Z}_{\text{t,t}} \vect{j}_{\text{t}} \\ \label{open_circuit_MIMO}
\vect{v}_{\text r} & = \vect{H} \vect{j}_{\text t},
\end{align}
with $\vect{H}$ the channel matrix.
The contribution to the transmit voltages of power reradiated back by the receive antennas is 
zero in \eqref{transmit_unilateral}; it
is negligible even with a finite-impedance receiver as per the \emph{unilateral approximation} \cite{Wallace2004,Nossek2010}, which rests on the strong signal attenuation at radio frequencies.

Communication theorists tend to ignore the coupling among transmit antennas by setting $\vect{Z}_{\text{t,t}} = \vect{I}_{N_\text{t}}$ 
in \eqref{transmit_unilateral}, whereby
the transmit circuit power is $P_{\text t} = \|\vect{j}_{\text t}\|^2$.
Together with \eqref{open_circuit_MIMO}, this forms the (noiseless) uncoupled MIMO channel model. While exceedingly useful, this model is not physical. 


\subsection{Continuous Multiport System Model}

Let $(\hat{\vect{x}}_\text{r},\hat{\vect{y}}_\text{r},\hat{\vect{z}}_\text{r})$ and $(\hat{\vect{x}}_\text{t},\hat{\vect{y}}_\text{t},\hat{\vect{z}}_\text{t})$ be orthonormal bases describing coordinate systems locally at receiver and transmitter, respectively. With respect to the former, a point in space is represented by the vector $r_x \hat{\vect{x}}_\text{r} + r_y \hat{\vect{y}}_\text{r} + r_z \hat{\vect{z}}_\text{r}$ with coordinates $\vect{{\sf r}} = (\vect{r},r_z)$ with $\vect{r} = (r_x,r_y)$; ditto with respect to the latter. 

Letting the number of transmit antennas grow unboundedly within a compact space, what arises is a continuum of ports described by the current and voltage space-time fields 
specified by their complex phasors $j(\vect{{\sf r}})$ and $v(\vect{{\sf r}})$.
The interactions between $\vect{{\sf r}}$ and some other point $\vect{{\sf s}}$ are described by a complex impedance kernel
\begin{equation}
z(\vect{{\sf r}},\vect{{\sf s}}) = \frac{v(\vect{{\sf r}})}{j(\vect{{\sf s}})} ,
\end{equation}
which is the continuous counterpart to \eqref{mutual_coupling_coeff}. 
We differentiate between the transmit and receive impedance kernels, $z_{\text{t,t}}(\vect{{\sf r}},\vect{{\sf s}})$ and $z_{\text{r,r}}(\vect{{\sf r}},\vect{{\sf s}})$, and the transimpedance kernel $h(\vect{{\sf r}},\vect{{\sf s}})$.

For continuous apertures,
\eqref{transmit_unilateral} morphs into the convolution
\begin{equation} \label{voltage}
v_{\text t}(\vect{{\sf s}}) = (\mathcal{Z}_{\text{t,t}} j_{\text{t}})(\vect{{\sf s}}) = \iiint_{-\infty}^\infty  z_{\text{t,t}}(\vect{{\sf s}},\vect{{\sf t}}) j_{\text t}(\vect{{\sf t}}) d\vect{{\sf t}}
\end{equation}
with $\mathcal{Z}_{\text{t,t}} : L^2 \to L^2$ the operator analogue to $\vect{Z}_{\text{t,t}} : \Complex^{N_\text{t}} \to \Complex^{N_\text{t}}$. This operator is nonconjugate and self-adjoint, meaning that $z(\vect{{\sf r}},\vect{{\sf s}}) = z(\vect{{\sf s}},\vect{{\sf r}})$, consistent with the physical reciprocity of the corresponding impedance matrices. 

The transimpedance, in turn, depends on the propagation medium and is generally not self-adjoint.
From \eqref{open_circuit_MIMO}, it can be regarded as the impulse response of the uncoupled MIMO channel, being
\begin{equation} \label{convolution}
v_{\text{r}}(\vect{{\sf r}}) = (\mathcal{H} j_{\text{t}})(\vect{{\sf r}}) 
= \iiint_{-\infty}^\infty h(\vect{{\sf r}},\vect{{\sf s}}) j_{\text{t}}(\vect{{\sf s}}) d\vect{{\sf s}}
\end{equation}
with $\mathcal{H} : L^2 \to L^2$ the 
continuous analogue to $\vect{H} : \Complex^{N_\text{t}} \to \Complex^{N_\text{r}}$. 
Ignoring coupling at the transmitter entails $z_{\text{t,t}}(\vect{{\sf r}})$ being impulsive in \eqref{voltage}, whereby the transmit circuit power is
\begin{equation} \label{uncoupled_power}
P_\text{t} 
= \iiint_{-\infty}^\infty |j_\text{t}(\vect{{\sf s}})|^2 d\vect{{\sf s}}.
\end{equation}

\section{Transmit Coupling Kernel \\ for Punctiform Antennas} \label{sec:coupling_iso}

For a lossless system, conservation of energy requires the transmit circuit power to equal the power
of the produced
electromagnetic field. 
These powers, derived in Appendix~A for punctiform antennas, are respectively given by 
\begin{align}    \label{circuit_power}
{\sf P}_{\text c}  & = \frac{1}{2} \Re\left\{\iiint_{-\infty}^\infty d\vect{{\sf r}} \, \overline{j_{\text t}(\vect{{\sf r}})} \iiint_{-\infty}^\infty d\vect{{\sf s}}  \, z_{\text{t,t}}(\vect{{\sf r}},\vect{{\sf s}}) j_{\text t}(\vect{{\sf s}}) \right\}
\end{align}
and
\begin{align}   \label{wave_power_final}
{\sf P}_\text{em}  & = \frac{1}{2} \Re\left\{ - \imagunit \kappa Z_0 \iiint_{-\infty}^\infty d\vect{{\sf r}} \,  \overline{j_{\text t}(\vect{{\sf r}})} \iiint_{-\infty}^\infty d\vect{{\sf s}} \, g(\vect{{\sf r}} - \vect{{\sf s}}) j_{\text t}(\vect{{\sf s}}) \right\}
\end{align}
where ${Z_0 \approx 120 \pi}$ is the wave impedance of free space while
\begin{equation} \label{Green}
g(\vect{{\sf r}}) = \frac{e^{\imagunit \kappa \|\vect{{\sf r}}\|}}{4\pi \|\vect{{\sf r}}\|}
\end{equation}
is the Green's function and
$\kappa = 2\pi/\lambda$ is the wavenumber, given $\lambda$ as the wavelength.
As they coincide, the two above powers are henceforth not distinguished, but
 unified into ${\sf P}_{\text t}$.

\subsection{Spectral Representation of the Transmit Impedance}

Define $\vect{{\sf v}}=(\vect{v},v_z)$, with $\vect{v}=\vect{r}-\vect{s}$ and $v_z=r_z-s_z$, as the space-lag variable.
The equality between the circuit power in \eqref{circuit_power} and the electromagnetic power in \eqref{wave_power_final} 
yields
\begin{align} \label{impedance_kernel_iso}  
z_{\text{t,t}}(\vect{{\sf v}}) & = -\imagunit \kappa Z_0 \,  g(\vect{{\sf v}})
= -\imagunit \kappa Z_0 \frac{e^{\imagunit \kappa \|\vect{{\sf v}}\|}}{4\pi \|\vect{{\sf v}}\|}
\end{align}
for $\|\vect{{\sf v}}\|>0$.
The transmit impedance kernel 
is space-invariant and takes the form of a spherical wave emanating from $\vect{{\sf s}}$ to any other point $\vect{{\sf r}}$.
 
An exact Fourier representation of the above impedance kernel 
is derived next, using Weyl's identity \cite{ChewBook}
\begin{equation} \label{Weyl}
\frac{e^{\imagunit \kappa \|\vect{{\sf v}}\|}}{\|\vect{{\sf v}}\|} =  \frac{\imagunit}{2\pi} \iint_{-\infty}^\infty  \frac{e^{\imagunit (\vect{\kappa}^{\Ttran} \vect{v} + \gamma |v_z|)}}{\gamma(\vect{\kappa})} d\vect{\kappa},
\end{equation} 
with $\gamma$ defined as
\begin{equation} \label{gamma}
\gamma(\vect{\kappa}) = 
\begin{cases}
\sqrt{\kappa^2 - \|\vect{\kappa}\|^2} & \quad \|\vect{\kappa}\|\le \kappa\\
\imagunit \sqrt{\|\vect{\kappa}\|^2 - \kappa^2} & \quad \|\vect{\kappa}\|> \kappa .
\end{cases}
\end{equation}
Plugging \eqref{Weyl} into \eqref{impedance_kernel_iso} yields 
\begin{align}  \label{impedance_kernel_spectral}
z_{\text{t,t}}(\vect{{\sf v}}) & =  \frac{\kappa Z_0}{2}
\iint_{-\infty}^\infty  \frac{e^{\imagunit \gamma |v_z|}}{\gamma(\vect{\kappa})} e^{\imagunit \vect{\kappa}^{\Ttran} \vect{v}}  \frac{d\vect{\kappa}}{(2\pi)^2}.
\end{align}
The spectrum contributing to the impedance is the one lying on a wavenumber hemisphere of radius $\kappa = 2\pi/\lambda$, either $\kappa_z=\gamma$ or $\kappa_z= - \gamma$ depending on the
location of $\vect{{\sf r}}$ relative to $\vect{{\sf s}}$. 

\subsection{Transmit Coupling Kernel}


Our attention now turns to the real part of the impedance, which is responsible for the transmit power. 
As shown in Appendix~B, \eqref{circuit_power} can be rewritten as 
\begin{align}  \label{transmit_power_zeta}
{\sf P}_\text{t} & = \frac{1}{2} \iiint_{-\infty}^\infty \! d\vect{{\sf r}} \, \overline{j_{\text{t}}(\vect{{\sf r}})} \iiint_{-\infty}^\infty \! d\vect{{\sf s}} \, \Re\{z_{\text{t,t}}(\vect{{\sf r}}-\vect{{\sf s}})\} j_{\text{t}}(\vect{{\sf s}}).
\end{align}
It is customary \cite{Wallace2004,Nossek2010}
to express the real part of $z_{\text{t,t}}(\vect{{\sf v}})$ as
\begin{equation} \label{normalization}
\Re\{z_{\text{t,t}}(\vect{{\sf v}})\} = {\sf R} \, {\sf c}_{\text t}(\vect{{\sf v}})
\end{equation}
where ${\sf c}_{\text t}(\vect{{\sf v}})$ is the transmit coupling kernel and ${\sf R} = \kappa^2 Z_0/4\pi$ is the radiation resistance, such that ${\sf c}_{\text t}(\vect{0}) = 1$.
With that, \eqref{transmit_power_zeta} becomes
\begin{align}  \label{transmit_power}
{\sf P}_\text{t} & = \frac{{\sf R}}{2} \iiint_{-\infty}^\infty \! d\vect{{\sf r}} \, \overline{j_{\text{t}}(\vect{{\sf r}})} \iiint_{-\infty}^\infty \! d\vect{{\sf s}} \, {\sf c}_\text{t}(\vect{{\sf r}}-\vect{{\sf s}}) j_{\text{t}}(\vect{{\sf s}}),
\end{align}
subsuming \eqref{uncoupled_power} for ${\sf c}_{\text t}(\vect{{\sf v}}) = \delta(\vect{{\sf v}})$, which embodies the special case of no coupling at the transmitter---a case that cannot arise from physical principles. 
Rather, the coupling with punctiform antennas is captured by a non-impulsive 
${\sf c}_\text{t}(\vect{{\sf r}})$. 
Precisely, applying Euler's formula to \eqref{impedance_kernel_iso},
\begin{align}  \label{real_impedance_kernel_spherical}
{\sf c}_{\text t}(\vect{{\sf v}}) = \sinc \! \left ( 2 \frac{\|\vect{{\sf v}} \|} {\lambda} \right) .
\end{align}

\begin{figure}
\centering\vspace{-0.0cm}
\includegraphics[width=.8\linewidth]{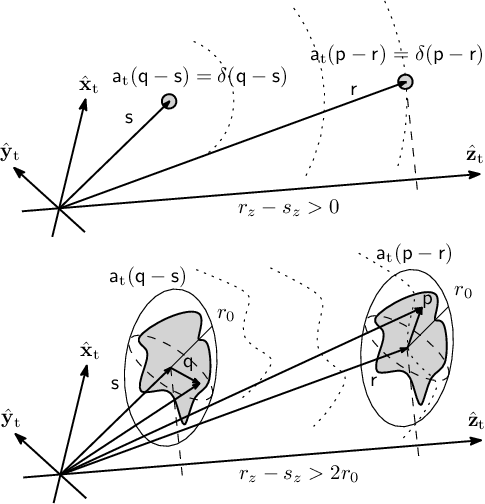} 
\caption{Mutual coupling between spatially causal antennas centered at $\vect{{\sf s}}$ and $\vect{{\sf r}}$. Top: the coupling between punctiform antennas takes the form of a spherical wave from $\vect{{\sf s}}$ to $\vect{{\sf r}}$. Bottom: physical antennas with arbitrary responses, ${\sf a}_\text{t}(\bm{\cdot})$; the coupling arises from the superposition of spherical waves emitted from points $\vect{{\sf p}}$ on the transmitting antenna's skin and received at points $\vect{{\sf q}}$ on the receiving antenna's skin.}\vspace{-0cm}
\label{fig:impedance_corr_tot}
\end{figure}

The discretization with punctiform antennas amounts to an ideal spatial sampling of the continuous current (see Fig.~\ref{fig:impedance_corr_tot}).
From \eqref{real_impedance_kernel_spherical}, antennas with half-wavelength spacing are uncoupled whereas, for arbitrary spacing, they are in general coupled.
An analogy can be established between \eqref{real_impedance_kernel_spherical} and
the autocorrelation of an isotropic random channel
\cite{teal2002spatial},
with uncoupled antennas being the analogue of antennas experiencing IID fading.
Indeed, as will be seen, coupling can be regarded as an additional spatial correlation---one that, unlike actual fading correlation, is not due to the angular spreading caused by scattering, but rather to the antenna structure. 
 
Momentarily disregarding the $z$-component, which amounts to a translation,
for $\|\vect{\kappa}\| > \kappa$ the spectrum in \eqref{impedance_kernel_spectral} is imaginary due to 
\eqref{gamma}. 
But the Fourier transform of \eqref{real_impedance_kernel_spherical}, which is  real and even,
must be real.
Thus, retaining only the portion $\|\vect{\kappa}\|\le \kappa$ of the impedance spectrum in \eqref{impedance_kernel_spectral}, under proper normalization, 
\begin{align}  \label{real_impedance_kernel}
{\sf c}_{\text t}(\vect{{\sf v}}) & =  
\frac{1}{2\pi \kappa} 
\iint_{\|\vect{\kappa}\|\le \kappa}  \frac{1}{\gamma(\vect{\kappa})} \, e^{\imagunit (\vect{\kappa}^{\Ttran} \vect{v} + \gamma |v_z|)}  d\vect{\kappa},
\end{align}
which implies the exclusion of evanescent waves; these do not contribute to the time-average transmit power
\cite{PlaneWaveBook}.
Consequently, the region of convergence of \eqref{real_impedance_kernel} extends to all $\vect{{\sf v}}$.
The above integration is intended on a $\kappa$-radius hemisphere, subject to the 2D parametrization
\begin{equation} \label{wavenumber_hemisphere}
(\vect{\kappa},\pm\gamma) : \{\|\vect{\kappa}\|\le \kappa\} \to \{\|\vect{\kappa}\|^2+\gamma^2=\kappa^2\}
\end{equation}
whose Jacobian, already reflected in \eqref{real_impedance_kernel}, is proportional to $\sqrt{\| \nabla \gamma\|^2 + 1}= 1/\gamma$ \cite{PizzoJSAC20}.
The upper and lower hemisphere supports in \eqref{real_impedance_kernel} respectively relate to the causal and anticausal parts of ${\sf c}_{\text t}(\vect{{\sf v}})$ along the $z$-axis.
The notion of causality applies here in the spatial domain and requires that $v_z>0$, which maps to $r_z > s_z$ $\forall s_z$; see Fig.~\ref{fig:impedance_corr_tot} (top). 

\subsection{Isotropic Coupling}

The 
coupling kernel
in \eqref{real_impedance_kernel_spherical} can be written as the Fourier transform
\begin{align}  \label{real_impedance_kernel_full}
{\sf c}_{\text t}(\vect{{\sf v}}) & = 
  \iiint_{-\infty}^\infty {\sf C}_{\text t}(\vect{{\sf k}}) \, e^{\imagunit \vect{{\sf k}}^{\Ttran} \vect{{\sf v}}} \frac{d\vect{{\sf k}}}{(2\pi)^3} 
\end{align}
of the spectrum
\begin{equation} \label{spectrum_impedance_isotropic}
{\sf C}_{\text t}(\vect{{\sf k}}) = \frac{4\pi^2}{\kappa} \, \delta \! \left(\|\vect{{\sf k}}\|^2 - \kappa^2 \right)
\end{equation}
given $\vect{{\sf k}} = (\vect{\kappa},\kappa_z) \in \Real^3$ the wavevector. The normalization ensures, as advanced, that ${\sf c}_{\text t}(\vect{{\sf 0}})=1$.
With punctiform antennas, therefore, a physically meaningful impedance must have 
a real part whose
spectrum lives on the skin of the wavenumber sphere as per \eqref{spectrum_impedance_isotropic}.
This is incompatible with any model ignoring mutual coupling, as a constant spectrum $\forall \vect{{\sf k}} \in \Real^3$ would be required for ${\sf c}_{\text t}(\vect{{\sf v}}) = \delta(\vect{{\sf v}})$.


Once more, an analogy can be drawn between mutual coupling and autocorrelation. 
Precisely, \eqref{spectrum_impedance_isotropic} is isomorphic with \cite[Eq.~27]{PizzoJSAC20}, the latter being the power spectral density of an 
isotropic random channel fading in 3D.

\section{Transmit Coupling Kernel \\ for Physical Antennas} \label{sec:coupling}


Realizable antennas are non-infinitesimal, which corresponds to sampling with a non-impulsive response \cite{Unser1994}. 
This is a 
real function 
that physically describes the antenna skin whereon currents may exist. It specifies the directivity pattern, with narrower patterns requiring a larger antenna structure as per the uncertainty principle.
We henceforth assume all antennas are identical with space-invariant response ${\sf a}_{\text t}(\vect{{\sf v}}) \in L^2$ 
and a corresponding 3D spectrum ${\sf A}_\text{t}(\vect{{\sf k}})$.
Then, the complex current density with physical antennas is specified by the convolution
\begin{align}  \label{current_LSI_filtering_discrete}
j^\prime_\text{t}(\vect{{\sf r}}) = \iiint_{-\infty}^\infty  j_\text{t}(\vect{{\sf s}}) \, {\sf a}_\text{t}(\vect{{\sf r}}-\vect{{\sf s}}) \, d\vect{{\sf s}}
\end{align}    
with $j_\text{t}(\vect{{\sf r}})$ the current corresponding to punctiform antennas.
Replacing the current in \eqref{wave_power_final} with \eqref{current_LSI_filtering_discrete} returns the transmit electromagnetic power 
\begin{align}  \nonumber
{\sf P}_\text{em}  & = \frac{1}{2} \Re \Big\{ - \imagunit \kappa Z_0  \iiint_{-\infty}^\infty d\vect{{\sf r}} \,  \overline{j_\text{t}(\vect{{\sf r}})}
\iiint_{-\infty}^\infty d\vect{{\sf s}} \, j_\text{t}(\vect{{\sf s}}) \\& \hspace{0.5cm}  \label{aa} 
\cdot \iiint_{-\infty}^\infty d\vect{{\sf p}} 
  \iiint_{-\infty}^\infty d\vect{{\sf q}}  \,    {\sf a}_\text{t}(\vect{{\sf p}}-\vect{{\sf r}}) g(\vect{{\sf p}} - \vect{{\sf q}}) {\sf a}_\text{t}(\vect{{\sf q}}-\vect{{\sf s}})   \Big\}.
\end{align}
Equated to the circuit power in \eqref{circuit_power}, 
the above yields the impedance kernel with physical antennas
\begin{align} \nonumber
z_{\text{t,t}}(\vect{{\sf v}})  & = -\imagunit \kappa Z_0  \iiint_{-\infty}^\infty d\vect{{\sf p}} \,  {\sf a}_\text{t}(\vect{{\sf p}}-\vect{{\sf r}})  \\& \hspace{1cm}  \label{bb} 
\cdot \iiint_{-\infty}^\infty d\vect{{\sf q}}  \,  g(\vect{{\sf p}}-\vect{{\sf q}}) {\sf a}_\text{t}(\vect{{\sf q}}+\vect{{\sf v}}-\vect{{\sf r}}),
\end{align}
as a function of the space lag, $\vect{{\sf v}} = \vect{{\sf r}} - \vect{{\sf s}}$.
This kernel, which reduces to \eqref{impedance_kernel_iso} for ${\sf a}_\text{t}(\vect{{\sf p}}) = \delta(\vect{{\sf p}})$,
 takes the form of a superposition of spherical waves between any two points $\vect{{\sf p}}$ and $\vect{{\sf q}}$ on the antenna skins, with centroids respectively at $\vect{{\sf r}}$ and $\vect{{\sf s}}$, as illustrated in Fig.~\ref{fig:impedance_corr_tot}.
Reciprocity holds in \eqref{bb} due to Green's function symmetry, $g(\vect{{\sf p}}-\vect{{\sf q}}) = g(\vect{{\sf q}} - \vect{{\sf p}})$. 

\subsection{Spectral Representation of the Transmit Impedance Kernel}

Consider an antenna physically confined to a sphere of radius $r_0\ge0$, as shown in Fig.~\ref{fig:impedance_corr_tot} (bottom).
Then, centroids of adjacent antennas must be separated by at least $2 r_0$ as antennas cannot overlap spatially.
This inherent causality and the antenna directionality are captured by the Fourier representation of the transmit impedance kernel, derived in Appendix~C as
\begin{align}    \nonumber
z_{\text{t,t}}(\vect{{\sf v}})  & =  \frac{\kappa Z_0}{2} 
\iint_{-\infty}^\infty \frac{d\vect{\kappa}}{(2\pi)^2} \,  \frac{e^{\imagunit \vect{\kappa}^{\Ttran} \vect{v}}}{\gamma(\vect{\kappa})}  \\ \label{impedance_kernel_spectral_noniso} 
& \hspace{2cm} \cdot
\begin{cases} \displaystyle
|{\sf A}_\text{t}^+(\vect{\kappa})|^2  \, e^{\imagunit \gamma v_z} & \quad v_z\ge0  \\ \displaystyle
|{\sf A}_\text{t}^-(\vect{\kappa})|^2  \, e^{-\imagunit \gamma v_z} & \quad v_z<0
\end{cases} 
\end{align} 
for $\|\vect{{\sf v}}\| > 2 r_0$, where ${\sf A}_\text{t}^\pm(\vect{\kappa}) = {\sf A}_\text{t}(\vect{\kappa},\pm \gamma)$ is the antenna pattern obtained evaluating the 3D spectrum at $\kappa_z = \pm \gamma$.
These patterns specify the non-isotropic coupling in the causal/anticausal direction.

For $r_0\to 0$, the antenna becomes infinitesimal and ${\sf a}_{\text t}(\vect{{\sf r}}) \to \delta(\vect{{\sf r}})$ whereby \eqref{impedance_kernel_spectral_noniso} correctly reduces to \eqref{impedance_kernel_spectral}. 
As in \eqref{impedance_kernel_spectral}, the symmetry of $z_{\text{t,t}}(\vect{{\sf v}})$ is established for any real-valued ${\sf a}_{\text t}(\vect{{\sf r}})$, as 
${\sf A}_{\text t}^\pm(-\vect{\kappa}) = \overline{{\sf A}^\pm_{\text t}(\vect{\kappa})}$ due to Hermitian symmetry.

\subsection{Non-Isotropic Coupling}

Excluding the evanescent portion of the spectrum from \eqref{impedance_kernel_spectral_noniso}, the Fourier representation in \eqref{real_impedance_kernel} generalizes to 
\begin{align}  \label{real_impedance_kernel_antenna}
{\sf c}_{\text t}(\vect{{\sf v}}) & =  
\frac{1}{2\pi \kappa} \iint_{\|\vect{\kappa}\|\le \kappa} \!\!\!\!\!\! d\vect{\kappa} \, \frac{e^{\imagunit \vect{\kappa}^{\Ttran} \vect{v}}}{\gamma(\vect{\kappa})} 
\cdot
\begin{cases} \displaystyle
|{\sf A}_{\text t}^+(\vect{\kappa})|^2 \, e^{\imagunit \gamma v_z}   & \; v_z\ge 0 \\\displaystyle
|{\sf A}_{\text t}^-(\vect{\kappa})|^2 \, e^{-\imagunit  \gamma v_z}  & \; v_z< 0
\end{cases} 
\end{align}
with 
\begin{align} \label{norm_A_spectrum}
1  & = \frac{1}{4 \pi \kappa}  \iint_{\|\vect{\kappa}\| \le \kappa} \frac{|{\sf A}_\text{t}^+(\vect{\kappa})|^2  + |{\sf A}_\text{t}^-(\vect{\kappa})|^2}{\gamma(\vect{\kappa})} \, d\vect{\kappa},
\end{align}
which translates the normalization ${\sf c}_{\text t}(\vect{{\sf 0}}) = 1$ to the spectral domain (see Appendix~D).
Invoking the identity
\begin{equation} \label{delta_nonlin}
\delta(\|\vect{{\sf k}}\|^2 - \kappa^2) = \frac{\delta(\kappa_z - \gamma) + \delta(\kappa_z + \gamma)}{2\gamma(\vect{\kappa})},
\end{equation}
the normalization \eqref{norm_A_spectrum} becomes $\iiint_{-\infty}^\infty {\sf C}_\text{t}(\vect{{\sf k}}) {d\vect{{\sf k}}}/{(2\pi)^3}  = 1$ with a spectrum
\begin{equation} \label{spectrum_impedance} 
{\sf C}_{\text t}(\vect{{\sf k}}) = \frac{4 \pi^2}{\kappa} \, |{\sf A}_{\text{t}}(\vect{{\sf k}})|^2 \, \delta(\|\vect{{\sf k}}\|^2 - \kappa^2),
\end{equation}
which generalizes \eqref{spectrum_impedance_isotropic} (i.e., ${\sf A}_{\text{t}}(\vect{{\sf k}})=1$) to 
physical antennas.
In the analogy between coupling and spatial correlation,
\eqref{spectrum_impedance} is isomorphic with the power spectral density of a stationary random channel fading \cite[Eq.~14]{PizzoJSAC20}.

\subsection{Transmit Power Spectral Density}


Plugging \eqref{real_impedance_kernel_antenna} into \eqref{transmit_power}, the transmit power becomes
\begin{align}  \nonumber
{\sf P}_\text{t} & = \frac{{\sf R}}{4\pi \kappa}
\iint_{\|\vect{\kappa}\|\le \kappa}  \!  \bigg(\frac{|J_{\text t}^+(\vect{\kappa})|^2 |{\sf A}_\text{t}^+(\vect{\kappa})|^2}{\gamma(\vect{\kappa})}  \\ \label{time_avg_power_final} 
& \hspace{3cm} 
+ \frac{|J_{\text t}^-(\vect{\kappa})|^2 |{\sf A}_\text{t}^-(\vect{\kappa})|^2}{\gamma(\vect{\kappa})}\bigg)  \, d\vect{\kappa}
\end{align}
where $J_{\text t}^\pm(\vect{\kappa}) = J_{\text t}(\vect{\kappa},\pm \gamma)$ with
\begin{equation} \label{3Dcurrent_spectrum}
J_{\text t}(\vect{{\sf k}}) = \iiint_{-\infty}^\infty  j_{\text t}(\vect{{\sf s}}) \, e^{-\imagunit \vect{{\sf k}}^{\Ttran} \vect{{\sf s}}} d\vect{{\sf s}}
\end{equation}
the 3D spectrum of the source.
Here, $|J_{\text t}^+(\vect{\kappa})|^2$ and $|J_{\text t}^-(\vect{\kappa})|^2$ are associated with the power flow along the $z$-axis, respectively causal and anticausal, measured on any $z$-plane.

By means of \eqref{delta_nonlin}, we can rewrite  \eqref{time_avg_power_final} as 
\begin{equation} \label{power_psd}
{\sf P}_\text{t} =  \iiint_{-\infty}^{\infty} {\sf S}_{\text t}(\vect{{\sf k}}) \frac{d\vect{{\sf k}}}{(2\pi)^3}
\end{equation}
given the 
power spectral density
\begin{align}  \label{psd_Dirac_LSI_noniso}
{\sf S}_{\text t}(\vect{{\sf k}}) & =  \frac{4 \pi^2 {\sf R}}{\kappa} \, |{\sf A}_{\text t}(\vect{{\sf k}})|^2 |J_{\text t}(\vect{{\sf k}})|^2 \delta(\|\vect{{\sf k}}\|^2 - \kappa^2)
 \\&  \label{psd_Dirac_LSI_noniso_filter}
= {\sf R} \, |J_{\text t}(\vect{{\sf k}})|^2 {\sf C}_\text{t}(\vect{{\sf k}}),
\end{align}
after substituting \eqref{spectrum_impedance}.
Rotationally symmetric sources, punctiform included, are indistinguishable in propagation as they generate the same power density.
Specifically, due to rotational symmetry, ${\sf a}_{\text t}(\vect{{\sf r}}) = {\sf a}_{\text t}(\|\vect{{\sf r}}\|)$, leading to ${\sf A}_{\text t}(\vect{{\sf k}}) = 1$ as per the wavenumber constraint in \eqref{wavenumber_hemisphere} 
and the normalization in \eqref{norm_A_spectrum}.

\section{MIMO Model with Transmit Coupling} \label{sec:MIMO_coupled}

\subsection{Composition of Channel and Coupling}

Let us rewrite \eqref{transmit_power} as 
\begin{align} \label{psd_op}
{\sf P}_\text{t} = \langle\mathcal{C} j_\text{t}, j_\text{t}\rangle 
\end{align}
with $\mathcal{C}$ the 
operator associated with 
the real 
kernel
$\frac{{\sf R}}{2} {\sf c}_{\text t}(\vect{{\sf v}})$ given ${\sf c}_{\text t}(\vect{{\sf v}})$ in \eqref{real_impedance_kernel_antenna}. 
By virtue of its positive-definiteness and symmetry, $(\mathcal{C} j_\text{t})(\vect{{\sf r}}) = (\mathcal{C}^{1/2} \mathcal{C}^{1/2} j_\text{t})(\vect{{\sf r}})$
for some other 
real symmetric and positive-definite operator $\mathcal{C}^{1/2}$ with associated kernel $\sqrt{{\sf R}/2} \,{\sf c}^{1/2}_{\text{t}}(\vect{{\sf v}})$, such that
 \begin{align} \label{conv_ct_3D}
{\sf c}_\text{t}(\vect{{\sf v}}) = \iiint_{-\infty}^\infty \!   {\sf c}^{1/2}_{\text{t}}(\vect{{\sf t}}) {\sf c}^{1/2}_{\text{t}}(\vect{{\sf v}} - \vect{{\sf t}}) \, d\vect{{\sf t}}.
\end{align}
Then,  \eqref{psd_op} can be rewritten as
\begin{align} \label{psd_op_continue}
{\sf P}_\text{t} & 
= \langle\mathcal{C}^{1/2} \mathcal{C}^{1/2} j_\text{t}, j_\text{t}\rangle \\  \label{Pt_inner_product_half}
&= \langle  \mathcal{C}^{1/2} j_\text{t},  \mathcal{C}^{1/2} j_\text{t}\rangle  \\
& = \| {\sf j}_\text{t} \|^2,
\end{align} 
where we exploited the symmetry of $\mathcal{C}^{1/2}$ and defined as
\begin{equation} \label{j_C_op} 
{\sf j}_\text{t}(\vect{{\sf s}}) = (\mathcal{C}^{1/2} j_\text{t})(\vect{{\sf s}}) = \sqrt{\frac{{\sf R}}{2}} \iiint_{-\infty}^\infty  {\sf c}^{1/2}_{\text{t}}(\vect{{\sf s}}-\vect{{\sf t}}) \, j_{\text{t}}(\vect{{\sf t}}) \, d\vect{{\sf t}}
\end{equation}
the composition of current density and coupling.
Rewriting \eqref{convolution} in terms of \eqref{j_C_op} leads to
\begin{align}  \label{MIMO_model_C_cont}
v_{\text{r}}(\vect{{\sf r}}) &
= ({\sf H} \, {\sf j}_\text{t})(\vect{{\sf r}})  
= \iiint_{-\infty}^\infty {\sf h}(\vect{{\sf r}},\vect{{\sf s}}) \, {\sf j}_\text{t}(\vect{{\sf s}}) \, d\vect{{\sf s}},
\end{align}
where ${\sf H} = \mathcal{H} \mathcal{C}^{-1/2}$ is the composition of the operator
modeling the uncoupled channel,
$\mathcal{H}$, with $\mathcal{C}^{-1/2}$, 
positive-definite and associated with the kernel $\sqrt{2/{\sf R}} \, {\sf c}^{-1/2}_{\text{t}}(\vect{{\sf v}})$.
The composition so defined associates with the space-variant kernel 
\begin{align}  
{\sf h}(\vect{{\sf r}},\vect{{\sf s}}) & = (\mathcal{C}^{-1/2} h)(\vect{{\sf r}},\vect{{\sf s}})  \\ \label{composite_channel_op}
& =  \sqrt{\frac{2}{{\sf R}}}\iiint_{-\infty}^\infty \!\!\! h(\vect{{\sf r}},\vect{{\sf t}}) \, {\sf c}^{-1/2}_{\text{t}}(\vect{{\sf t}}-\vect{{\sf s}}) \, d\vect{{\sf t}}.
\end{align}

 The invertibility of $\mathcal{C}^{1/2}$ is proven in Appendix~E;
 it requires that $|{\sf A}_{\text{t}}(\vect{{\sf k}})|$ be strictly positive almost everywhere.

The model in \eqref{MIMO_model_C_cont} is equivalent to that of \eqref{convolution} in the sense that both provide the same output,
with ${\sf j}_\text{t}(\vect{{\sf r}})$ and ${\sf h}(\vect{{\sf r}},\vect{{\sf s}})$ providing an alternative embodiment of the MIMO formulation, in some ways advantageous relative to $j_{\text{t}}(\vect{{\sf s}})$ and $h(\vect{{\sf r}},\vect{{\sf s}})$.

\subsection{Composite MIMO Channel Matrix}

Before proceeding, let us verify that sampling our continuous model at antenna locations yields established results for discrete arrays \cite{Nossek2010}. 
Sampling the convolution in \eqref{MIMO_model_C_cont} yields the noiseless MIMO relationship inclusive of transmit coupling 
  \begin{equation} \label{convolution_coupling_sampled}
   \vect{v}_\text{r} = \vect{{\sf H}} \,\vect{{\sf j}}_\text{t},
    \end{equation}
where $[\vect{{\sf H}}]_{n,m} = {\sf h}(\vect{{\sf r}}_n,\vect{{\sf s}}_m)$ and $[\vect{{\sf j}}_\text{t}]_m = {\sf j}_\text{t}(\vect{{\sf s}}_m)$ are the composite channel matrix and composite current vector, respectively. 
These are obtainable from \eqref{composite_channel_op} and \eqref{j_C_op}  as 
    \begin{align} \label{channel_samples_coupled}
\vect{{\sf H}} = \sqrt{\frac{2}{{\sf R}}} \, \vect{H} \vect{{\sf C}}_\text{t}^{-1/2}
\end{align}
and 
    \begin{align} \label{composite_current_vec}
\vect{{\sf j}}_\text{t} = \sqrt{\frac{{\sf R}}{2}} \, \vect{{\sf C}}_\text{t}^{1/2} \vect{j}_\text{t}.
\end{align}
Expanding the space-lag variable in \eqref{conv_ct_3D} as $\vect{{\sf v}} = \vect{{\sf r}} - \vect{{\sf s}}$ and applying the change of variables $\vect{{\sf r}}-\vect{{\sf t}} = \vect{{\sf y}}$, 
 \begin{align}  \label{conv_ct_expanded}
{\sf c}_\text{t}(\vect{{\sf r}}-\vect{{\sf s}}) 
& = \iiint_{-\infty}^\infty \! {\sf c}^{1/2}_{\text{t}}(\vect{{\sf r}}-\vect{{\sf y}}) {\sf c}^{1/2}_{\text{t}}(\vect{{\sf y}}-\vect{{\sf s}}) \, d\vect{{\sf y}}.
\end{align} 
Sampling \eqref{conv_ct_expanded} at any two transmit antenna locations and exploiting the matrix symmetry yields
$[\vect{{\sf C}}_\text{t}]_{n,m} = {\sf c}_\text{t}(\vect{{\sf r}}_n-\vect{{\sf s}}_{m})$.     
The composite current and composite channel coincide with the 
    circuit-theoretic formulas \cite[Eqs.~99, 101]{Nossek2010} specialized to the multiport settings of Fig.~\ref{fig:multiport} and uncorrelated noise.
The lack of reciprocity in \eqref{channel_samples_coupled}, compared to \cite[Eq.~101]{Nossek2010}, arises from the absence of coupling among receive antennas, as no current is drawn from them (see Fig.~\ref{fig:multiport}). 
In turn, the factor $1/2$ arises from the passband formulation, in contrast with the baseband formulation in \cite{Nossek2010};
the baseband bandwidth is half the passband bandwidth.

\subsection{Transmit Coupling Matrix}

%
%

Changing the domain from wavenumber to spherical, it is found that
\begin{equation} \label{wavenumber_spherical}
\vect{\kappa}(\theta,\phi) = \left(\kappa \sin \theta \cos \phi, \kappa \sin \theta \sin \phi \right),
\end{equation}
and $\gamma(\theta) = \sqrt{\kappa^2 - \|\vect{\kappa}\|^2} = \kappa \cos \theta$, with Jacobian 
\begin{equation} \label{Jacobian}
\frac{\partial \vect{\kappa}}{ \partial(\theta,\phi)}  = \kappa^2 \cos \theta \sin \theta.
\end{equation}
With these variable changes applied to \eqref{real_impedance_kernel_antenna}, 
\begin{align} \notag
[\vect{{\sf C}}_\text{t}]_{n,m} & =  
\frac{1}{4\pi}
\int_0^\pi \int_0^{2\pi}  |{\sf A}_\text{t}(\theta_\text{t},\phi_\text{t})|^2 \\ & \hspace{1cm} \label{impedance_matrix_general} 
 \cdot a_n(\theta_{\text t},\phi_{\text t}) \overline{a_{m}(\theta_{\text t},\phi_{\text t})}
 \sin \theta_{\text t}  \,  d\theta_{\text t} \, d\phi_{\text t},
\end{align}
where the array response vector is
\begin{align}  \label{Gavi}
a_{n}(\theta_{\text t},\phi_{\text t}) & = e^{\imagunit (\vect{\kappa}^{\Ttran}(\theta_\text{t},\phi_\text{t}) \vect{r}_n + \gamma(\theta_\text{t}) r_{z,n})}
\end{align}
with $\vect{\kappa}(\cdot,\cdot)$ as per \eqref{wavenumber_spherical} and with
${\sf A}_\text{t}(\cdot,\cdot)$ being either ${\sf A}_\text{t}^+(\cdot,\cdot)$ over the upper hemisphere, $\{\theta_\text{t} \in [0,\frac{\pi}{2}]\}$,
 or ${\sf A}_\text{t}^-(\cdot,\cdot)$ over the lower hemisphere, $\{\theta_\text{t} \in (\frac{\pi}{2}, \pi]\}$.
From \eqref{norm_A_spectrum},
\begin{align} \label{norm_A_spectrum_angle}
1  & = \frac{1}{4\pi} \int_0^\pi \int_0^{2\pi}  |{\sf A}_\text{t}(\theta_{\text t},\phi_{\text t})|^2 \, \sin \theta_{\text t}  \,  d\theta_{\text t} \, d\phi_{\text t}
\end{align}
with omnidirectional coupling, including punctiform antennas, arising for ${\sf A}_\text{t}(\theta_\text{t},\phi_\text{t}) = 1$.  
The function $|{\sf A}_\text{t}(\theta_{\text t},\phi_{\text t})|^2$ is the 
 power pattern of each antenna. 

For a $z$-aligned array of punctiform antennas, 
\eqref{impedance_matrix_general} reduces to the impedance formula in \cite[Eq.~44]{Nossek2010}.
And, for a uniform linear array (ULA) in particular,
\begin{align}  \label{Toeplitz_coupling}
[\vect{{\sf C}}_\text{t}]_{n,m}
= \sinc \! \left (2 \frac{\|\vect{{\sf r}}_n-\vect{{\sf s}}_{m}\|} {\lambda} \right),
\end{align}
consistent with \eqref{real_impedance_kernel_spherical}.
For a ULA, a real-symmetric Toeplitz structure emerges for the transmit coupling matrix while, for a uniform planar array (UPA), it is block-Toeplitz. 
In general, uncoupled MIMO would require 
$\vect{{\sf C}}_\text{t} = \vect{I}_{N_\text{t}}$, meaning transmit antennas that are infinitely apart from each other.
This is incongruent with the transmitter fitting in a certain form factor. 
    
\section{Fourier Model for MIMO Channels \\ Without Coupling} \label{sec:holo_MIMO_uncoupled}

\subsection{Fourier Representation of Stationary Channels}


The channel kernel in \eqref{convolution} comprises 
a causal
and 
 an anticausal
component at each end of the link, resulting in four possible combinations. 
The 
 causal-causal component of a stationary Gaussian channel 
can be represented in Fourier form as \cite[Thm.~2]{PizzoIT21}
\begin{align}  \notag
h(\vect{{\sf r}},\vect{{\sf s}}) & = \iint_{\|\vect{k}\|\le\kappa} \frac{d\vect{k}}{\sqrt{2\pi \kappa}} \iint_{\|\vect{\kappa}\|\le\kappa} \frac{d\vect{\kappa}}{\sqrt{2\pi \kappa}} \, \frac{\tilde{H}^{++}(\vect{k},\vect{\kappa})}{\sqrt{\gamma(\vect{k}) \gamma(\vect{\kappa})}}  \\ & \hspace{2cm} \label{spectral_representation_z} 
 \cdot e^{\imagunit (\vect{k}^{\Ttran} \vect{r} + \gamma(\vect{k}) r_z)}  e^{-\imagunit (\vect{\kappa}^{\Ttran} \vect{s} + \gamma(\vect{\kappa}) s_z)}
\end{align}
with $\gamma(\vect{\cdot})$ in \eqref{gamma} and with $\tilde{H}^{++}(\vect{k},\vect{\kappa})$ the 
spectrum,  
an independent complex Gaussian process.
The dependences on $r_z$ and $s_z$ are henceforth omitted as they can be absorbed into $\tilde{H}^{++}(\vect{k},\vect{\kappa})$ and are immaterial due to the statistical equivalence 
at different $z$-planes, namely
\begin{equation}
\tilde{H}^{++}(\vect{k},\vect{\kappa}) \sim e^{\imagunit \gamma(\vect{k}) r_z} \tilde{H}^{++}(\vect{k},\vect{\kappa}) \, e^{-\imagunit \gamma(\vect{\kappa}) s_z}.
\end{equation}
We adhere to a separable model satisfying 
\begin{equation} \label{separable_corr}
\Ex\{|\tilde{H}^{++}(\vect{k},\vect{\kappa})|^2\} = \Ex\{|\tilde{H}^{+}_\text{r}(\vect{k})|^2\}
 \, \Ex\{|\tilde{H}^{+}_\text{t}(\vect{\kappa})|^2\},
\end{equation}
with $\Ex\{|\tilde{H}^{+}_\text{r}(\vect{k})|^2\}$ and $\Ex\{|\tilde{H}^{+}_\text{t}(\vect{\kappa})|^2\}$ the power angle spectra due to the separate scattering at receiver and transmitter, respectively.
These spectra are normalized so each can be interpreted as an angular distribution. For example,
\begin{align} \label{normalization_channel}
1 = \frac{1}{4\pi} \int_0^{\pi} \int_0^{2\pi} \Ex\{|\tilde{H}^{+}_\text{t}(\theta_{\text t},\phi_{\text t})|^2\}  \sin \theta_{\text t} \, d\theta_{\text t} d\phi_{\text t},
\end{align}
at the transmitter, with the variables change according to \eqref{wavenumber_spherical}.
This normalization 
bounds the channel power, scaling with the product of the transmit and receive apertures rather than the antenna count, in the limit of antenna densification \cite{PizzoTWC21}.

\subsection{Karhunen–Loève Decomposition}


The spatial limitation imposed by the array apertures implies that $h(\vect{{ r}},\vect{{ s}})$ can be approximated by a finite number of coefficients representing the channel's spatial dimensionality 
at either end of the link. 
For the sake of specificity, rectangular apertures of unnormalized dimensions $(L_{{\text t},x},L_{{\text t},y})$ at the transmitter and $(L_{{\text r},x},L_{{\text r},y})$ at the receiver are considered.
For later convenience, the shortest dimension aligns with the $x$-axis, whereby $L_{{\text r},x} \le L_{{\text r},y}$ and $L_{{\text t},x} \le L_{{\text t},y}$. 
Define 
\begin{equation} \label{n_rt}
 {\sf n}_{\text r} = 
 \lceil \pi \det(\vect{D}_\text{r}) \rceil
 \quad \qquad 
 {\sf n}_{\text t} = 
 \lceil \pi \det(\vect{D}_\text{t}) \rceil
\end{equation}
with $\vect{D}_\text{r} = \diag(L_{{\text r},x}, L_{{\text r},y})/\lambda$ and $\vect{D}_\text{t} = \diag(L_{{\text t},x}, L_{{\text t},y})/\lambda$.
The number of spatial dimensions or degrees of freedom (DOF) 
is upper bounded by its value under isotropic scattering, given by ${\sf DOF} = \min({\sf DOF}_\text{r},{\sf DOF}_\text{t})$, with \cite{PizzoTWC21}
\begin{equation} \label{DOF}
{\sf DOF}_{\text r}  = {\sf n}_{\text r} +  o \! \left({\sf n}_{\text r}\right)  \qquad 
{\sf DOF}_{\text t} = {\sf n}_{\text t} +  o \! \left({\sf n}_{\text t}\right)  
  \end{equation}
the cardinalities of the 2D lattices
\begin{align}    \label{lattice_rx}
\Lambda_{\text r} & = \left\{\vect{i} =(i_x,i_y) \in \Integer^2 : \|\vect{D}_\text{r}^{-1} \vect{i}\| \le 1\right\} \\   \label{lattice_tx}
\Lambda_{\text t} & = \left\{\vect{j} =(j_x,j_y) \in \Integer^2 : \|\vect{D}_\text{t}^{-1} \vect{j}\| \le 1\right\}.
\end{align} 

The MIMO channel $[\vect{H}]_{n,m} = h(\vect{{ r}}_n,\vect{{ s}}_m)$ can then be approximated by the Karhunen–Lo\`eve expansion \cite{PizzoTWC21,LozanoCorrelation}
\begin{align} \label{Kronecker_MIMO}
\vect{H} & \approx   
\vect{U} \tilde{\vect{H}} \vect{V}^{\Htran}
\end{align} 
where $\vect{U} \in \Complex^{N_\text{r} \times {\sf n}_\text{r}}$ and $\vect{V} \in \Complex^{N_\text{t} \times {\sf n}_\text{t}}$ are the isometry Fourier eigenvector matrices with column entries 
\begin{align} \label{u_vec}
[\vect{u}_\vect{i}]_n & =   \sqrt{\frac{L_{{\text r},x} L_{{\text r},y}}{N_\text{r}}} u_\vect{i}(\vect{r}_n)   
= \frac{1}{\sqrt{N_\text{r}}}  e^{\imagunit 2\pi (\vect{D}_\text{r}^{-1} \boldsymbol{i})^{\Ttran} \vect{r}_n/\lambda}   \\ \label{v_vec}
[\vect{v}_\vect{j}]_m & =    \sqrt{\frac{L_{{\text t},x} L_{{\text t},y}}{N_\text{t}}} v_\vect{j}(\vect{s}_m)  
  = \frac{1}{\sqrt{N_\text{t}}} e^{\imagunit 2\pi (\vect{D}_\text{t}^{-1} \vect{j})^{\Ttran} \vect{s}_m/\lambda}.
\end{align}
In turn, 
\begin{equation} \label{equiv_channel} 
\vect{\tilde{H}} = \vect{\Lambda}^{1/2}_{\text{r}} \vect{W} \vect{\Lambda}_{\text{t}}^{1/2}
\end{equation}
where $\vect{W} \in \Complex^{{\sf n}_\text{r} \times {\sf n}_\text{t}}$ has IID standard complex Gaussian entries while $\vect{\Lambda}_{\text{r}}$ and $\vect{\Lambda}_{\text{t}}$ are diagonal with entries that depend on the 
power spectra at receiver and transmitter, respectively.
For example, at the transmitter, $[\vect{\Lambda}_{\text{t}}]_{j,j} = N_\text{t} \sigma^2_{\vect{j}}(\tilde{H}^{+}_\text{t})$ for $j = 1, \ldots, {\sf n}_{\text{t}}$ an arbitrary ordering of the lattice points within $\Lambda_{\text t}$ in \eqref{lattice_tx} \cite{PizzoTWC21}
\begin{align}  \label{variances_channel}
\sigma^2_{\vect{j}}(\tilde{H}^{+}_\text{t}) & 
 = \frac{1}{2\pi}  \iint_{\Omega^+_{\vect{j}}}  \Ex\{|\tilde{H}^{+}_\text{t}(\theta_{\text t},\phi_{\text t})|^2\}  \sin \theta_{\text t} \, d\theta_{\text t} d\phi_{\text t}
\end{align} 
given $\Omega^+_{\vect{j}}$ the spherical surface elements covering the entire upper hemisphere, each one centered at
\begin{align} \label{midpoint}
\theta_{{\text t},\vect{j}} & = \sin^{-1} \! \left(\|\vect{D}_\text{t}^{-1} \vect{j}\| \right)  \qquad \phi_{{\text t},\vect{j}} = \tan^{-1} \! \left({j_y}/{j_x}\right).
\end{align}
Under isotropic scattering, $\Ex\{|\tilde{H}^{+}_\text{t}(\theta_{\text t},\phi_{\text t})|^2\} = 1$, and \eqref{variances_channel} reduces to the solid angle subtended by $\Omega^+_{\vect{j}}$,
 \begin{align}  \label{solid_angles}
|\Omega^+_{\vect{j}}| & 
 = \frac{1}{2\pi}  \iint_{\Omega^+_{\vect{j}}} \sin \theta_{\text t} \, d\theta_{\text t} \, d\phi_{\text t}.
\end{align}
With that, $\vect{\tilde{H}}$ specifies the angular coupling between resolvable transmit and receive directions, while the left and right multiplications by $\vect{U}$ and $\vect{V}$ change the domain from angular to spatial \cite{PizzoTWC21,Sayeed2002}.

As the apertures grow electrically large, the approximation in \eqref{Kronecker_MIMO} sharpens 
and, by virtue of Mercer's theorem, converges in the mean-square sense to \eqref{spectral_representation_z}. Concurrently, the lower-order terms in  \eqref{DOF} vanish as the number of DOF firms up.

\subsection{Model Accuracy with Finite Apertures}

\begin{figure}
\centering\vspace{-0.0cm}
\includegraphics[width=.999\linewidth]{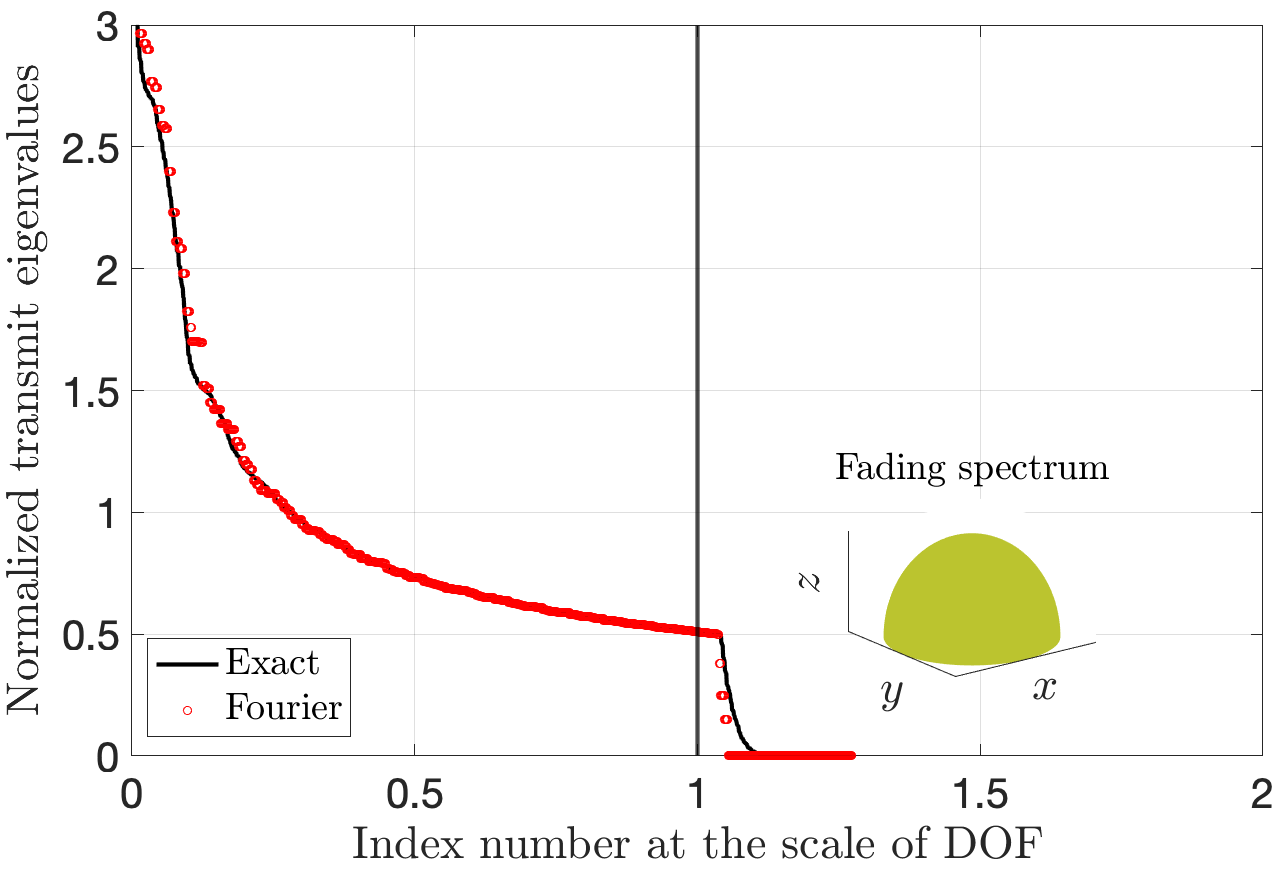} 
\caption{Normalized sorted eigenvalues of the transmit correlation matrix in an isotropic channel. UPA with uncoupled antennas spaced by $\lambda/2$ and aperture $20 \lambda$. The solid line indicates the exact channel in \eqref{corr_exact}, circles denote its Fourier approximation in \eqref{tx_corr}.}
\label{fig:channel_var_uncoupled}
\end{figure}

Denote the correlation between the $(n,m)$th and $(n^\prime,m^\prime)$th channel entries by \cite{LozanoCorrelation}
\begin{align} \label{corr}
R_\vect{H}(n,m;n^\prime,m^\prime) = \Ex\{[\vect{H}]_{n,m} [\vect{H}]^*_{n^\prime,m^\prime}\}.
\end{align}
Under the separable
model, \eqref{corr} reduces to the product of the marginal correlations at each end of the link, namely $R_\vect{H}(n,m;n^\prime,m^\prime) = [\vect{R}_\text{r}]_{n,n^\prime} \, [\vect{R}_\text{t}]_{m,m^\prime}$ with $\vect{R}_\text{r}$ and $\vect{R}_\text{t}$ the receive and transmit correlations, e.g.,
\begin{align} \label{tx_corr}
\vect{R}_\text{t} & = \frac{1}{N_\text{r}} \Ex\{\vect{H}^{\Htran} \vect{H}\} \approx \vect{V} \vect{\Lambda}_{\text{t}} \vect{V}^{\Htran}
\end{align}
at the transmitter, with $\vect{V}$ and $\vect{\Lambda}_\text{t}$ from 
\eqref{Kronecker_MIMO} and \eqref{equiv_channel}, respectively.
Holographic MIMO is characterized by \cite{PizzoTWC21}
\begin{equation} \label{Nyquist_cond}
N_{\text r} \ge {\sf n}_\text{r} \qquad \text{and} \qquad N_{\text t} \ge {\sf n}_\text{t}
\end{equation}
implying that $\vect{H}$ is rank-deficient with probability $1$, namely $\rank(\vect{H}) \le \min({\sf n}_\text{r},{\sf n}_\text{t}) \le \min(N_\text{r},N_\text{t})$, 
the number of spatial dimensions being limited by the environment and array apertures, rather than by the number of antennas as in regular MIMO \cite{heath_lozano_2018}.
Then, augmenting $\vect{V}$ with its orthogonal complement $\vect{V}^\perp \in \Complex^{N_\text{t} \times N_\text{t}-{\sf n}_\text{t}}$ gives the unitary matrix $[\vect{V},\vect{V}^\perp]$ whereby 
the eigenvalues of the transmit correlation read 
\begin{align} \label{tx_eig_uncoupled}
\vect{\lambda}(\vect{R}_\text{t}) \approx N_{\text t} \, \big(\underbrace{\sigma^2_{1}(\tilde{H}^{+}_\text{t}), \ldots, \sigma^2_{{\sf n}_\text{t}}(\tilde{H}^{+}_\text{t})}_\text{DOF}, \underbrace{0, \ldots, 0}_{N_\text{t}-{\sf n}_\text{t}} \big).
\end{align}
Fig.~\ref{fig:channel_var_uncoupled} depicts, at the scale of ${\sf n}_\text{t}$ in \eqref{n_rt}, the sorted eigenvalues \eqref{tx_eig_uncoupled} specialized for a squared UPA of uncoupled antennas spaced by $\lambda/2$ and aperture $L_{\text{t},x} = L_{\text{t},y} = 20 \lambda$.  
The curve is gauged against the eigenvalues of the exact correlation matrix, 
whose entries are obtained from \eqref{spectral_representation_z} with \eqref{separable_corr} and the variable changes in \eqref{wavenumber_spherical} as
\begin{align}  \notag
[\vect{R}_\text{t}]_{m,m^\prime} & = \frac{1}{2\pi}  \int_0^{\pi/2} \int_0^{2 \pi}  \Ex\{|\tilde{H}^{+}_\text{t}(\theta_{\text t},\phi_{\text t})|^2\}  \\& \label{corr_exact} \hspace{1cm}
\cdot
 a_{m}(\theta_{\text t},\phi_{\text t}) \overline{a_{m^\prime}(\theta_{\text t},\phi_{\text t})} \sin \theta_{\text t} \, d\theta_{\text t} d\phi_{\text t},
\end{align}
where $a_{m}(\cdot,\cdot)$ is the array response in \eqref{Gavi}.
Isotropic scattering is considered, implying 
that \eqref{corr_exact} specializes to $[\vect{R}_\text{t}]_{m,m^\prime} = \sinc(2 \|\vect{s}_m-\vect{s}_{m^\prime}\|/\lambda)$ and the variance in \eqref{tx_eig_uncoupled} reduces to the solid angle in \eqref{solid_angles}.
The eigenvalues exhibit a polarization at ${\sf n}_\text{t}$, justifying a low-rank representation via Fourier expansion, 
 with lower-order dependencies of the rank subsumed by the term $o({\sf n}_\text{t})$ in \eqref{DOF}.

\section{Fourier Model with Coupling at the Transmitter} \label{sec:holo_MIMO_coupled}


Electromagnetic propagation in 3D is representable in a 
2D form \cite{Franceschetti}.
This inherent lower dimensionality is revealed by a 2D Fourier transform of ${\sf c}_{\text t}(\vect{{\sf v}})$ in \eqref{real_impedance_kernel_antenna}, with $v_z$ kept fixed,
\begin{align}  
{\sf C}_{\text t}(\vect{\kappa},v_z) &  =  \iint_{-\infty}^\infty  {\sf c}_{\text t}(\vect{v},v_z) \, e^{-\imagunit \vect{\kappa}^{\Ttran} \vect{v} } d\vect{v} \\&  \label{real_impedance_kernel_antenna_spectrum}
= \frac{1}{2\pi \kappa} \mathbbm{1}_{\|\vect{\kappa}\|\le \kappa} \frac{|{\sf A}_{\text t}^+(\vect{\kappa})|^2}{\gamma(\vect{\kappa})} \, e^{\imagunit \gamma v_z},
\end{align}
which captures the spectral behavior of coupling between $z$-planes separated by $v_z\ge0$.
This spectrum evolves along the arbitrarily chosen $z$-axis as ${\sf C}_{\text t}(\vect{\kappa},v_z) = {\sf C}_{\text t}(\vect{\kappa},0) \, e^{\imagunit \gamma v_z}$, in the half-space $v_z \ge 0$.
The limitation to $\|\vect{\kappa}\|\le \kappa$ ensures that $\gamma \in \Real$, making the above relation equivalent to a 
 2D spectrum representation with translation along $z$.
While a 3D formulation 
 revealed the isomorphism between coupling and correlation in \eqref{spectrum_impedance_isotropic} and \eqref{spectrum_impedance}, 
the framework is henceforth specialized to planar apertures, whereby ${\sf C}_{\text t}(\vect{\kappa}) = {\sf C}_{\text t}(\vect{\kappa},0)$. (Volumetric apertures can be accommodated by superimposing contributions from current densities across different $z$-planes, leveraging the linearity of wave propagation.)

\subsection{Coupling as a Linear Space-Invariant Filter}

Replacing $h(\vect{{ r}},\vect{{ s}})$ with its spectral representation in \eqref{spectral_representation_z}, with the dependences on $r_z$ and $s_z$ omitted, while building on the space invariance of ${\sf c}^{-1/2}_{\text{t}}(\vect{{ r}}-\vect{{ s}})$, the composite channel ${\sf h}(\vect{{ r}},\vect{{ s}})$ in \eqref{composite_channel_op} between continuous apertures is obtained analogously to \eqref{spectral_representation_z}, but with a coupling-inclusive spectrum that, scaled by $\sqrt{2/{\sf R}}$, equals
\begin{align}  \label{spectrum_HC}
\tilde{{\sf H}}^{++}(\vect{k},\vect{\kappa}) & =  
 \tilde{H}^{++}(\vect{k},\vect{\kappa}) \,
 {\sf C}^{-1/2}_{\text{t}}(\vect{\kappa})
  \end{align}
 with 
\begin{align}   \label{C_square_root_inv}
{\sf C}^{-1/2}_{\text{t}}(\vect{k}) & = \frac{1}{\sqrt{{\sf C}_{\text{t}}(\vect{k})}} = 
\frac{\sqrt{\gamma(\vect{\kappa})}}{|{\sf A}_{\text t}^+(\vect{\kappa})|} \frac{\sqrt{2\pi \kappa}}{\mathbbm{1}_{\|\vect{k}\|\le\kappa}},
\end{align}
obtainable specializing \eqref{conv_ct_3D} to 2D apertures and substituting \eqref{real_impedance_kernel_antenna_spectrum}.
The invertibility of $\mathcal{C}^{1/2}$, ensuring the existence of a deconvolution $\mathcal{C}^{-1/2}$, requires the magnitude of the antenna pattern $|{\sf A}_{\text{t}}^+(\vect{\kappa})|$ to be strictly positive almost everywhere (see Appendix~E).
The stability of \eqref{C_square_root_inv} is further guaranteed for any bounded antenna pattern \cite{Unser1994}.

In light of \eqref{spectrum_HC}, the sole effect of coupling is to alter the spectrum of the channel via a space-invariant filter---akin to the separable correlation \cite{chizhik2000effect}.
From \eqref{spectrum_HC}, 
\begin{align} \label{spectrum_channel_transmitter_coupling}
\Ex \! \left \{ |\tilde{{\sf H}}^{+}_\text{t}(\vect{\kappa})|^2 \right \} 
= 2 \pi \kappa \gamma(\vect{\kappa}) \, \frac{\Ex  \big \{|\tilde{H}^{+}_\text{t}(\vect{\kappa})|^2 \big \}}{|{\sf A}_{\text t}^+(\vect{\kappa})|^2}  \mathbbm{1}_{\|\vect{k}\|\le\kappa},
\end{align}
where the channel spectrum's support in \eqref{spectral_representation_z} was made explicit through an indicator function.
This support excludes evanescent waves, which would violate the stationarity assumption in \eqref{spectral_representation_z} that underlies the Karhunen-Loève expansions, and carry no real power, as evidenced by \eqref{time_avg_power_final}.

\subsection{Coupling versus Correlation} \label{sec:coupling_vs_corr}

Despite fading correlation and coupling having an isomorphic linear filtering nature, their effects appear respectively in the numerator and denominator of \eqref{spectrum_channel_transmitter_coupling}. 
Correlation is associated with scattering and maps directly onto the channel, while coupling arises from the interactions among currents flowing into different antennas. 
Putting them on an equal footing requires translating coupling to the channel 
through the reciprocal 
in \eqref{C_square_root_inv}. 
This entails a deconvolution, as opposed to the convolution applied by scattering; while their functional dependence is isomorphic, therefore,
coupling effects run counter to those of scattering.

\subsection{Composite MIMO Channel Matrix}

The separable MIMO channel, including 
 transmit coupling, admits the approximate Karhunen–Lo\`eve expansion 
\begin{align} \label{MIMO_channel_coupling}
\vect{{\sf H}} & \approx   
 \sqrt{\frac{2}{{\sf R}}} \, \vect{U} \vect{\tilde{{\sf H}}} \vect{V}^{\Htran}
\end{align}
where $\vect{U}$ and $\vect{V}$ are defined in \eqref{Kronecker_MIMO}. In turn, 
\begin{equation} \label{equiv_channel_coupling} 
\vect{\tilde{{\sf H}}} = \vect{\Lambda}^{1/2}_{\text{r}} \vect{W} \vect{{\sf \Lambda}}_{\text{t}}^{1/2}
\end{equation}
where $\vect{{\sf \Lambda}}_{\text{t}}$ is diagonal with entries $\{N_\text{t} \sigma^2_{j}(\tilde{{\sf H}}_\text{t}^+)\}$.
These variances are obtained similarly to \eqref{variances_channel}, but with the coupling-inclusive spectrum, yielding
\begin{align} \label{variances_channel_composite}
\sigma^2_{\vect{j}}(\tilde{{\sf H}}_\text{t}^+)   
= \iint_{\Omega^+_{\vect{j}}}  \frac{\Ex\{|\tilde{H}^+_\text{t}(\theta_{\text t},\phi_{\text t})|^2\}}{|{\sf A}_\text{t}^+(\theta_{\text t},\phi_{\text t})|^2} \,\cos\theta_{\text t} \, \sin \theta_{\text t} \, d\theta_{\text t} d\phi_{\text t}
\end{align}
with the variable changes in  \eqref{wavenumber_spherical}.
Comparing \eqref{variances_channel_composite} against the variances of the uncoupled model in \eqref{variances_channel}, the antenna  power pattern appears at the denominator of the composite angular spectrum and a projection arises w.r.t. the broadside direction, orthogonal to the array plane. 
This projection emanates from coupling through the deconvolution applied by \eqref{C_square_root_inv} and incorporates the impact of array orientation for a given scattering.
For example, specialized to isotropic scattering and omnidirectional antennas, \eqref{variances_channel_composite} becomes (up to a factor) the projected solid angle subtended by $\Omega^+_{\vect{j}}$ in \eqref{variances_channel}, namely
\begin{equation} \label{solid_angles_coupling}
|{\sf \Omega}^+_{\vect{j}}| = \frac{1}{\pi} \iint_{\Omega^+_{\vect{j}}}  \cos\theta_{\text t} \, \sin \theta_{\text t} \, d\theta_{\text t} d\phi_{\text t}.
\end{equation}
Coupling introduces a broadside projection, absent in the uncoupled formulation---recall \eqref{solid_angles}. It incorporates the impact of array orientation, skewing the solid angles subtended by the scattering towards the broadside direction.

\subsection{Impact of Coupling on the Transmit Eigenvalues}  \label{sec:tx_eig}

\begin{figure}[t]
     \centering
     \includegraphics[width=.999\columnwidth]{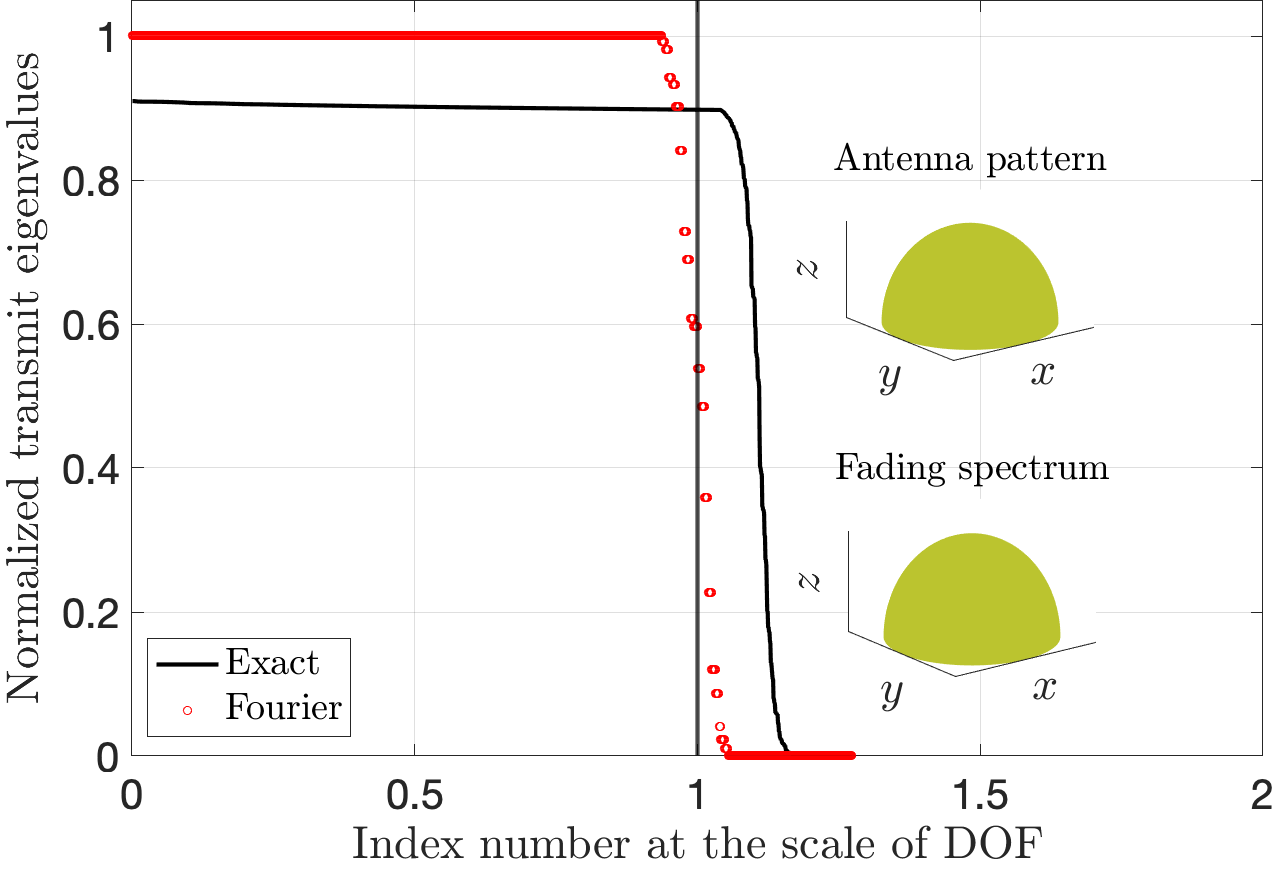}
         \caption{Normalized sorted eigenvalues of the transmit correlation matrix under isotropic scattering. UPA with omnidirectional antennas ($\rho = 0.01$) spaced by $\lambda/2$ and aperture $20 \lambda$. The solid line is for the exact channel in \eqref{numerical_inv}, circles for its Fourier approximation in \eqref{tx_corr_coupled}.
         }
     \label{fig:channel_var_coupled_iso}
     \end{figure}

Generalization of the transmit correlation matrix in \eqref{tx_corr} to coupled antennas yields
\begin{align}  \label{tx_corr_coupled}
\vect{{\sf R}}_\text{t} & = \frac{1}{N_\text{r}} \Ex\{\vect{{\sf H}}^{\Htran} \vect{{\sf H}}\}   \approx \frac{2}{{\sf R}} \,  \vect{V} \vect{{\sf \Lambda}}_{\text{t}} \vect{V}^{\Htran} 
\end{align}
with associated eigenvalues
\begin{align} \label{eig_tx_corr_coupled}
\vect{\lambda}(\vect{{\sf R}}_\text{t}) \approx  N_\text{t} \, \big(\underbrace{\sigma^2_{1}(\tilde{{\sf H}}^{+}_\text{t}), \ldots, \sigma^2_{{\sf n}_\text{t}}(\tilde{{\sf H}}^{+}_\text{t})}_\text{DOF}, \underbrace{0, \ldots, 0}_{N_\text{t}-{\sf n}_\text{t}} \big),
\end{align}
where $N_\text{t}-{\sf n}_\text{t}$ eigenvalues are zero, as in \eqref{tx_eig_uncoupled}, due to the inherent low-rankness of the Fourier representation.

When the antenna power pattern matches the fading spectrum, coupling 
causes antenna decorrelation, resulting in the most uniform distribution and strongest polarization 
of the eigenvalues.
This is because fading maps to the channel via convolution, whereas coupling acts as a deconvolution 
through the reciprocal of the pattern, as discussed in Sec.~\ref{sec:coupling_vs_corr}.
This stronger eigenvalue polarization is evidenced by the Fourier curve in Fig.~\ref{fig:channel_var_coupled_iso}, where the sorted eigenvalues in \eqref{eig_tx_corr_coupled} are plotted for isotropic scattering and omnidirectional antennas, whereby \eqref{variances_channel_composite} reduces to the projected solid angles in \eqref{solid_angles_coupling}.

Also shown in Fig.~\ref{fig:channel_var_coupled_iso} are the exact eigenvalues of the transmit correlation, under the same normalization, derivable from \eqref{channel_samples_coupled} as
\begin{align} \label{numerical_inv}
\vect{{\sf R}}_\text{t} = \frac{2}{{\sf R}} \, \vect{{\sf C}}^{-1/2}_\text{t} \vect{R}_\text{t} \vect{{\sf C}}^{-1/2}_\text{t}
\end{align}
with $\vect{R}_\text{t}$ the correlation without coupling, given in \eqref{corr_exact}.
The potential for coupling to decorrelate the antennas can also be appreciated here. Ultimately,
if $\vect{{\sf C}}_\text{t}$ equals $\vect{R}_\text{t}$, then \eqref{numerical_inv} becomes an identity matrix.
However, accounting for ohmic losses in the antennas, a portion of the transmit power dissipates as heat, subsumed by the augmented matrix \cite{Nossek2010}
\begin{align} \label{coupling_loss}
\vect{{\sf C}}_\text{t}(\rho) = \vect{{\sf C}}_\text{t} + \rho \vect{I}_{N_\text{t}},
\end{align}
with $0 < \rho < 1$ the loss factor.
This loss is tantamount to a physical regularization, improving the conditioning of $\vect{{\sf C}}_\text{t}$ for a stable inversion, but at the cost of a reduced decorrelation.
\section{DOF Augmentation via Mutual Coupling} \label{sec:DOF_increase}

\begin{figure}[t!]
     \centering
     \begin{subfigure}{\columnwidth}
         \centering
         \includegraphics[width=.9\columnwidth]{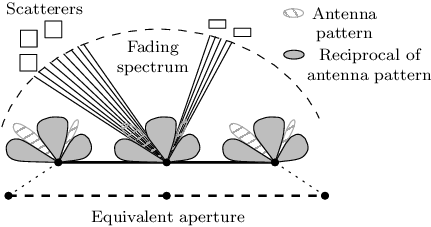}
\caption{Reciprocal of the antenna pattern runs counter to the fading spectrum.}
         \label{fig:decorrelation_2a}
     \end{subfigure}
     \vfill
     \vspace{.2cm}
     \begin{subfigure}{\columnwidth}
         \centering
         \includegraphics[width=.9\columnwidth]{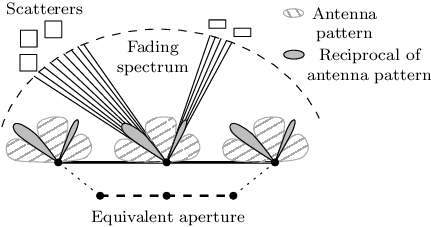}
         \caption{Reciprocal of the antenna pattern matches the fading spectrum.}
         \label{fig:decorrelation_2b}
     \end{subfigure}
       \caption{{\color{blue}Having the reciprocal of the antenna power pattern counter or match the fading spectrum causes coupling to 
       reduce or enhance antenna correlation, respectively. This is because fading maps to the channel via
convolution in \eqref{spectrum_HC}, whereas coupling acts as a deconvolution through the reciprocal of the pattern.}
       For a fixed antenna count, reducing correlation appears to expand the antenna spacing, hence the aperture, while increasing correlation is interchangeable with compressing the antenna spacing, hence the aperture.}
        \label{fig:decorrelation_2}
\end{figure}

\subsection{Impact of Antenna Power Pattern on DOF}

The exact eigenvalues in Fig.~\ref{fig:channel_var_coupled_iso} reveal a second effect that the Fourier approximation fails to capture, namely that,
properly harnessed, coupling can create \emph{additional} DOF 
beyond the uncoupled limit at ${\sf n}_{\text t}$. 
An interpretation of this phenomenon is provided in Fig.~\ref{fig:decorrelation_2}, whereby---everything else being the same---a reduced correlation is indistinguishable from a wider antenna spacing, hence an enlarged aperture for a fixed antenna count.
Conversely, an increased correlation 
is tantamount to having tighter antenna spacings, making the aperture appear smaller for a fixed antenna count.
Thus, consistent with Landau's theorem, which expresses the spatial DOF as the product of spatial bandwidth and aperture \cite{Franceschetti,PizzoWCL22}, extra DOF may arise due to coupling: more or fewer DOF fit on the equivalent aperture, as it is expanded or contracted by coupling.
Such DOF augmentation is reflected in the horizontal stretch of the exact curve in Fig.~\ref{fig:channel_var_coupled_iso}, and also its vertical shift (due to the normalization by a larger number of positive eigenvalues.)

\subsection{Impact of Antenna Losses and Densification on DOF}

\begin{figure}[t]
     \centering
     \includegraphics[width=.999\columnwidth]{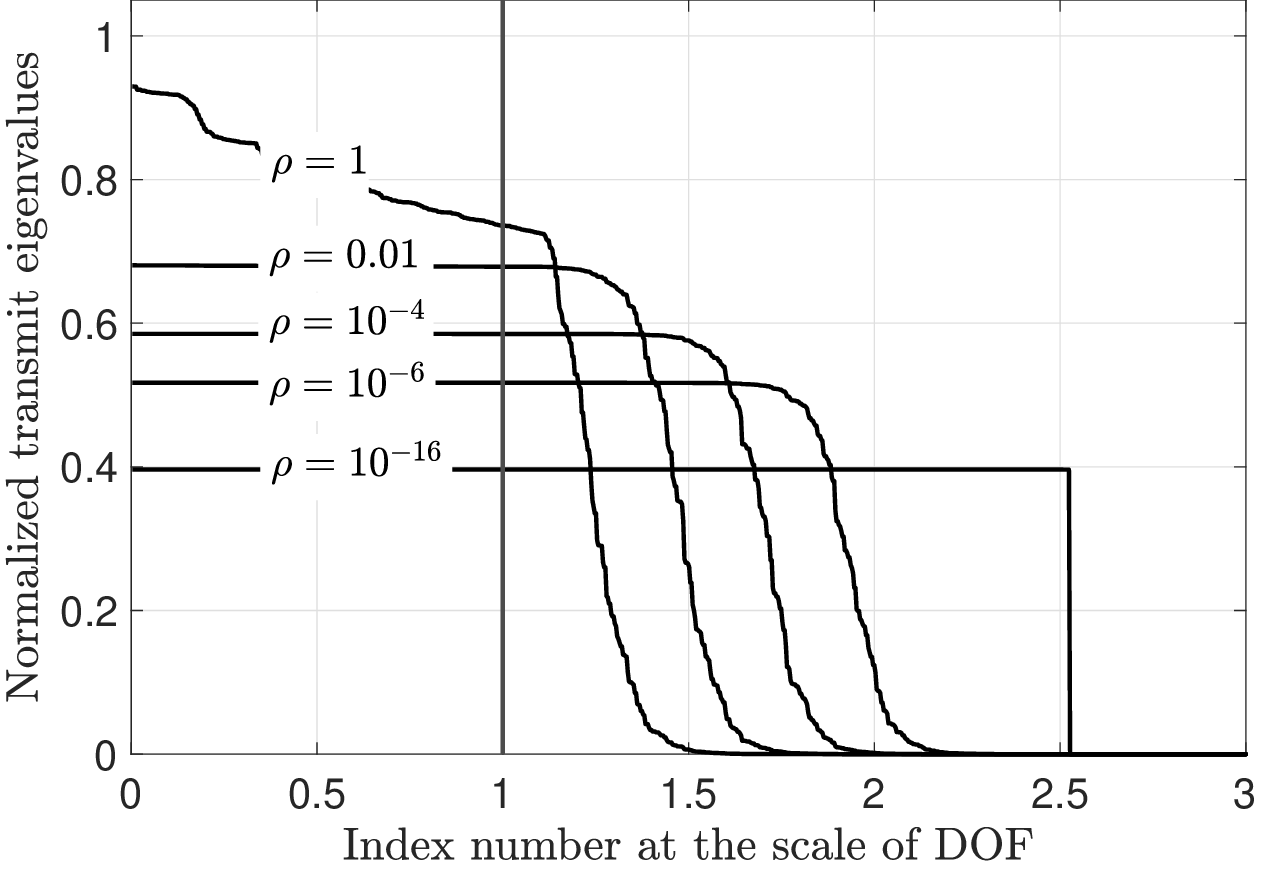}
         \caption{Spatial DOF augmentation for various $\rho$ under isotropic scattering for the exact correlation in \eqref{numerical_inv}. Omnidirectional antennas (decorrelating the fading), 
         spaced by $\lambda/4$ and aperture $10 \lambda$.}
     \label{fig:DOF_augmentation}
     \end{figure}

Antenna losses (in the amount of $\rho$, the loss factor) tone down the impact of the antenna pattern as a proxy for transmit coupling, limiting the extent to which the equivalent aperture expands or contracts for a fixed antenna count, with $\rho=0$ and $\rho=1$ the most and least favorable to DOF augmentation, respectively.
The sorted eigenvalues in Fig.~\ref{fig:DOF_augmentation} illustrate this for various $\rho$ and a fixed aperture, $L_{\text{t},x}=L_{\text{t},y}=10 \lambda$. Antennas are omnidirectional, matching the considered isotropic scattering.
The extra DOF brought about by coupling can be leveraged if the antenna density is sufficiently high, up to what the sampling theorem determines for the enlarged aperture. The Nyquist density thus depends on $\rho$ and on the fading spectrum. For $\rho\to 0$, each additional antenna should give rise to an equal increase in DOF, as hinted by the inversion in \eqref{numerical_inv}. However, due to finite accuracy of computer simulations, this limit is not achievable in numerical simulations.
Indeed, there is probably no limit to the theoretical DOF gain that can be achieved by such ideal antennas, but this result is of limited significance when physical antennas are concerned---antenna size constraint and losses limit the achievable gain.

\section{Information-Theoretic Analysis}   \label{sec:spectral_efficiency}
  
  
Let us now turn to the information-theoretic implications of mutual coupling at different SNR levels. To isolate its effect on the eigenvalue polarization from that of having extra DOF, the latter aspect is excluded in this section and the Fourier model is applied unless otherwise stated. The impact of the additional DOF could be separately studied by replacing the actual aperture by its coupling-expanded counterpart.
  
  
 \subsection{Coupling as Physical Precoder}

Communicating entails encoding independent unit-variance symbols $\vect{s} \in\Complex^{{\sf n}_{\text t}}$ onto the current vector $\vect{j}_{\text t} = \vect{F} \vect{s} $ via the 
precoder $\vect{F} \in\Complex^{N_{\text t} \times {\sf n}_{\text t}}$.
From \eqref{convolution_coupling_sampled} and \eqref{composite_current_vec}, the input-output relationship over a noisy channel is thus
\begin{equation} \label{MIMO_coupling}
\vect{y} = \vect{{\sf H}} \vect{{\sf F}} \vect{s} + \vect{n} 
\end{equation}
with $\vect{{\sf H}}$ the composite MIMO channel and $\vect{n}\sim\CN(\vect{0}, \sigma^2 \vect{I}_{N_{\text r}})$ given $\sigma^2$ the noise power, which could be spatially colored if interference were present in addition to thermal noise. 
In turn,
\begin{equation} \label{FC_precoder}
\vect{{\sf F}} = \vect{{\sf C}}_\text{t}^{1/2} \vect{F} \in \Complex^{N_{\text t} \times {\sf n}_{\text t}}
\end{equation}
is the composition of precoder and coupling, subject to
\begin{equation} \label{SNR_constraint}
{\rm tr} \! \left(\vect{{\sf F}}^{\Htran} \vect{{\sf F}}\right) \le  {\sf SNR}
\end{equation}
where ${\sf SNR} = \frac{2}{{\sf R}} \frac{G {\sf P}_\text{t}}{\sigma^2}$ is the average 
signal-to-noise-ratio,
with $G$ the large-scale channel gain. 
The above SNR constraint 
prevents the radiated power from growing with either $N_{\text t}$ or with ${\sf n}_{\text t}$, i.e., with the number of transmit antennas or the transmit aperture \cite{Wallace2004}. 

The effect of the mutual coupling is seen to be that of a precoder that, in tandem with the standard precoder, linearly combines the entries of the input vector. 
Unlike that standard precoder, though, the coupling precoder is induced by antenna interactions.
This is in agreement with previous works on mutual coupling \cite{Nossek2010,Wallace2004}.

\subsection{Ergodic Capacity with Coupling at the Transmitter}


With perfect channel-state information (CSI) at the receiver,
the mutual information 
achieved by $\vect{s}\sim\CN(\vect{0},\vect{I}_{{\sf n}_{\text t}})$
is \cite{heath_lozano_2018} 
\begin{align}  \label{MI_xy_composite}
I(\vect{s};\vect{y}| \vect{{\sf H}}) & 
= \log_2 \det ( \vect{I} + \vect{{\sf H}} \vect{{\sf F}} \vect{{\sf F}}^{\Htran} \vect{{\sf H}}^{\Htran} ) .
\end{align}  
The rotational invariance of $\vect{s}$ renders
 the right singular vectors of the precoder immaterial, whereby 
\begin{equation} \label{precoder_coupling}
\vect{{\sf F}} = \vect{U}_{\vect{{\sf F}}} \vect{{\sf P}}^{1/2}
\end{equation}
with $\vect{U}_{\vect{{\sf F}}}  \in\Complex^{N_{\text t} \times {\sf n}_{\text t}}$ 
and $\vect{{\sf P}} = \diag({\sf p}_1, \ldots, {\sf p}_{{\sf n}_{\text t}})$, ${\sf p}_i \ge 0$ such that \eqref{SNR_constraint} is satisfied, implying $\sum_{i=1}^{{\sf n}_{\text t}} {\sf p}_i \le {\sf SNR}$.

The superscript $^\star$ is henceforth used to distinguish the capacity-achieving value of any quantity. 
With CSI further at the transmitter, \eqref{MI_xy_composite} is maximized when
$\vect{U}_{\vect{{\sf F}}}^\star = \vect{V}_{\vect{{\sf H}}}$ \cite[Sec. 5.3]{heath_lozano_2018} with $\vect{V}_{\vect{{\sf H}}}$ the right singular vector matrix of $\vect{{\sf H}}$, 
\begin{equation} \label{right_singular_tildeHc} 
\vect{V}_{\vect{{\sf H}}} \approx \vect{V} \vect{V}_{\vect{\tilde{{\sf H}}}} \in \Complex^{N_\text{t}\times {\sf n}_\text{t}}
\end{equation}
with $\vect{V}$ the 
Fourier matrix in \eqref{Kronecker_MIMO} and $\vect{V}_{\vect{\tilde{{\sf H}}}} \in \Complex^{{\sf n}_\text{t}\times {\sf n}_\text{t}}$ the right singular vector matrix of $\vect{\tilde{{\sf H}}}$ in \eqref{equiv_channel_coupling}.
By exploiting the Fourier matrix unitarity, we deviate from the usual MIMO formulation, which expresses the precoder in terms of $\vect{V}_{\vect{{\sf H}}}$ directly. 


The ergodic capacity for a specific fading spectrum and antenna pattern
is then 
\begin{align} \label{capacity}
{\sf C}({\sf SNR})  =  \Ex \! \left\{ \sum_{i=1}^{{\sf n}_\text{min}} \log_2 \! \left(1 + {\sf p}_i^\star \lambda_{i}(\vect{\tilde{{\sf H}}} \vect{\tilde{{\sf H}}}^{\Htran}) \right)\right\}
\end{align}
where 
\begin{equation} \label{waterfilling}
{\sf p}_i^\star = \left(\nu - \lambda_{i}^{-1}(\vect{\tilde{{\sf H}}} \vect{\tilde{{\sf H}}}^{\Htran})\right)^{\! +}
\end{equation}
with ${\sf n}_\text{min} = \min({\sf n}_\text{r},{\sf n}_\text{t})$ and $\nu$ such that $\sum_{i=1}^{{\sf n}_\text{min}} {\sf p}_i^\star = {\sf SNR}$, given $\lambda_{i}(\vect{\tilde{{\sf H}}} \vect{\tilde{{\sf H}}}^{\Htran})$ as the $i$th unordered eigenvalue of $\vect{\tilde{{\sf H}}} \vect{\tilde{{\sf H}}}^{\Htran}$. 
The transmit precoder achieving \eqref{capacity} arises from \eqref{FC_precoder} after substituting \eqref{precoder_coupling} and solving for $\vect{F}$,
\begin{align}  \label{optimal_precoder_0}
\vect{F}^\star & =  \vect{{\sf C}}_\text{t}^{-1/2} \vect{U}_{\vect{{\sf F}}}^\star (\vect{{\sf P}}^\star)^{1/2} \\ \label{optimal_precoder}
& = \vect{{\sf C}}_\text{t}^{-1/2} \vect{V} \vect{V}_{\vect{\tilde{{\sf H}}}}
(\vect{{\sf P}}^\star)^{1/2}
\end{align}
where \eqref{right_singular_tildeHc} was used.
In the special case that the channel is  line-of-sight, $\vect{H} = \vect{a}(\theta_{\text t},\phi_{\text t}) \vect{a}^{\Htran}(\theta_{\text t},\phi_{\text t})$ with $\vect{a}(\cdot,\cdot)$ the array response in \eqref{Gavi}, 
the right singular vector of $\vect{H}^{\Htran} \vect{H}$ is---recalling \eqref{channel_samples_coupled}---proportional to $\vect{{\sf C}}_\text{t}^{-1/2} \vect{a}(\theta_{\text t},\phi_{\text t})$, whereby 
\cite{Marzetta2019}
\begin{align}
\vect{F}^\star = \sqrt{{\sf SNR}} \, \vect{{\sf C}}_\text{t}^{-1} \vect{a}(\theta_{\text t},\phi_{\text t}).
\end{align}

A precoder that ignored coupling would convey information spatially on a set of resolvable directions determined by the scattering environment and specified by the right singular vectors of 
$\vect{H}^{\Htran} \vect{H}$, namely \cite{PizzoTWC21}
 \begin{equation}
 \vect{F}^\star =  \vect{V} \vect{V}_{\vect{\tilde{H}}} (\vect{P}^\star)^{1/2}
 \end{equation}
where $\vect{P}^\star$ and $\vect{V}_{\vect{\tilde{H}}}$ are associated with $\vect{\tilde{H}}$ in \eqref{equiv_channel}, rather than $\vect{\tilde{{\sf H}}}$ in \eqref{equiv_channel_coupling}.
Instead, the optimal precoder in \eqref{optimal_precoder} first diagonalizes $\vect{\tilde{{\sf H}}}$, which does depend on the coupling. The result is translated to the spatial domain via $\vect{V}$
and then whitened to remove the coupling produced by proximal transmit antennas.

\subsection{Low-SNR Regime}  


At low SNR, the precoder allocates the entire power budget to the maximal-eigenvalue eigenspace of $\tilde{\vect{{\sf H}}} \tilde{\vect{{\sf H}}}^{\Htran}$, yielding an equivalent single-antenna channel with signal-to-noise $\lambda_\text{max}(\tilde{\vect{{\sf H}}} \tilde{\vect{{\sf H}}}^{\Htran}) {\sf SNR}$.
Thus, \eqref{capacity} expands as \cite[Eq. 5.38]{heath_lozano_2018}
\begin{align}   \label{lowSNR_capacity}
{\sf C}({\sf SNR}) & =  \Ex\{\lambda_\text{max}(\tilde{\vect{{\sf H}}} \tilde{\vect{{\sf H}}}^{\Htran})\} \, {\sf SNR} 
+ o({\sf SNR}).
\end{align}


With a view to drawing insights, we resort to the bound derived in Appendix~F, which separates the effect of coupling from the channel characteristics and array geometry, 
namely
\begin{align}  \label{bound_eig_HC_discrete}  
\Ex \! \left\{ \lambda_\text{max}\big(\tilde{\vect{{\sf H}}} \tilde{\vect{{\sf H}}}^{\Htran}\big) \right\}  & \le
\underbrace{N_{\text r} N_{\text t}}_\text{Array gain} \,  
\underbrace{\mathop{\max}\limits_{\vect{i} \in \Lambda_\text{r}} 
 \sigma^2_{\vect{i}}(\tilde{{H}}_\text{r}^+)}_\text{Receive Correlation}
  \\ & \quad \notag
\cdot 
\underbrace{\Ex \! \left\{\lambda_\text{max}(\vect{W} \vect{W}^{\Htran}) \right\}}_\text{Fading} 
\!\!\!\!\!\!
\underbrace{\mathop{\max}\limits_{\vect{j} \in \Lambda_\text{t}} \sigma^2_{\vect{j}}(\tilde{{\sf H}}_\text{t}^+).}_\text{Transmit Correlation/Coupling}
 \end{align}
Note that $ \sigma^2_{\vect{i}}(\tilde{{H}}_\text{r}^+)$ and $\sigma^2_{\vect{j}}(\tilde{{\sf H}}_\text{t}^+)$ are inversely proportional to the respective electrical apertures,
which could suggest that \eqref{bound_eig_HC_discrete} vanishes for an infinite aperture.
However, the number of antennas per dimension increases with the aperture as per \eqref{Nyquist_cond}, ensuring that 
the products $N_{\text r} \sigma^2_{\vect{i}}(\tilde{{H}}_\text{r}^+)$ and $N_\text{t} \sigma^2_{\vect{j}}(\tilde{{\sf H}}_\text{t}^+)$ remain finite asymptotically.

The first term  in \eqref{bound_eig_HC_discrete}
is the array gain 
with uncoupled omnidirectional  antennas, 
$\lambda_\text{max}\big(\tilde{\vect{H}} \tilde{\vect{H}}^{\Htran}\big) = \|\tilde{\vect{H}}\|_{\text F}^2 =  N_\text{r} N_{\text t}$.
The remaining terms in \eqref{bound_eig_HC_discrete} incorporate the influence of 
 fading correlation and transmit coupling. Some considerations:

\begin{itemize}
\item
Expectedly, stronger correlation is beneficial at low SNR for a given number of uncoupled antennas, as preferred directions arise that beamforming can exploit \cite{heath_lozano_2018}.
For large electrical apertures and a smooth fading spectrum,
\begin{align} \label{sigma2_asympt}
\sigma^2_{\vect{j}}(\tilde{{H}}_\text{t}^+) \approx |{\Omega}^+_{\vect{j}}| \, \Ex\{|H^{+}_\text{t}(\theta_{{\text t},\vect{j}},\phi_{{\text t},\vect{j}})|^2\}
\end{align}
after applying the midpoint integration rule to \eqref{variances_channel}, with \eqref{midpoint} and \eqref{solid_angles} substituted. Correlation increases \eqref{sigma2_asympt} and, in turn, \eqref{bound_eig_HC_discrete} as per the normalization in \eqref{normalization_channel}.
\item
Correlation can be further enhanced by leveraging coupling.
From \eqref{variances_channel_composite}, specifically, 
\begin{align} \label{sigma2_coupling_asympt}
\sigma^2_{\vect{j}}(\tilde{{\sf H}}_\text{t}^+) \approx |{\sf \Omega}^+_{\vect{j}}| \frac{\Ex\{|H^{+}_\text{t}(\theta_{{\text t},\vect{j}},\phi_{{\text t},\vect{j}})|^2\}}{|A^{+}_\text{t}(\theta_{{\text t},\vect{j}},\phi_{{\text t},\vect{j}})|^2} ,
\end{align}
which is maximized when the antenna pattern is lowest on directions on which the fading is strongest (see Fig.~\ref{fig:decorrelation_2b}).
\item
Coupling can alternatively decorrelate antennas, diminishing the selectivity (see Figs.~\ref{fig:channel_var_coupled_iso} and~\ref{fig:decorrelation_2a}).
The minimum of \eqref{sigma2_coupling_asympt} is attained by antenna patterns closely matching the fading spectrum, resulting in $\sigma^2_{\vect{j}}(\tilde{{\sf H}}_\text{t}^+) \approx |{\sf \Omega}^+_{\vect{j}}|$, which may be lower than the uncoupled value in \eqref{sigma2_asympt}.
\item
The highest value in \eqref{sigma2_coupling_asympt} is dictated by the maximum selectivity. 
For a fixed aperture, a trade-off arises between array gain, increasing with antenna densification, and antenna selectivity, requiring larger structures as dictated by the uncertainty principle.
\item
As the number of antennas grows with the electrical aperture,
a plural multiplicity of $\lambda_\text{max}(\tilde{\vect{{\sf H}}} \tilde{\vect{{\sf H}}}^{\Htran})$ arises due to the eigenvalue polarization \cite{Franceschetti,PizzoWCL22,HeedongIRS}. Then, low-SNR optimality entails multiple equal-power transmissions on each of those maximal-eigenvalue  eigenvectors.
\end{itemize}


 
\subsection{High-SNR Regime} \label{sec:DOF_HighSNR}




At high SNR, with probability 1 \cite[Eq. 5.29]{heath_lozano_2018}, 
\begin{align}  \label{C_SNR_high}
{\sf C}({\sf SNR}) & =  {\sf DOF} \, \log_2 {\sf SNR} + \mathcal{O}(1) ,
\end{align}
 where 
(see Appendix~G) 
\begin{equation} \label{S_inf_UIU_rank}
{\sf DOF} = \min({\sf DOF}_\text{r}^\prime,{\sf DOF}_\text{t}^\prime), 
\end{equation}
with ${\sf DOF}_{\text r}^\prime$ and ${\sf DOF}_{\text t}^\prime$ as defined in \eqref{DOF}, but with ${\sf n}_\text{r}^\prime \le {\sf n}_\text{r}$ and ${\sf n}_\text{t}^\prime \le {\sf n}_\text{t}$ incorporating the effects of fading and coupling via
\begin{align}   \label{n_r_prime}
{\sf n}_\text{r}^\prime & =  \left\lceil {\sf n}_\text{r}  \cdot \frac{1}{\pi} \!\iint_{{\rm supp}(\Ex\{|\tilde{H}^+_\text{r}|^2\})} \!\!\!  \cos \theta_\text{r} \, \sin \theta_\text{r} \, d\theta_\text{r} \, d\phi_\text{r} \right\rceil \\  \label{n_t_prime}
{\sf n}_\text{t}^\prime & =  \left\lceil {\sf n}_\text{t}  \cdot \frac{1}{\pi} \!\iint_{{\rm supp}(\Ex\{|\tilde{{\sf H}}^+_\text{t}|^2\})} \!\!\!  \cos \theta_\text{t} \, \sin \theta_\text{t} \, d\theta_\text{t} \, d\phi_\text{t} \right\rceil.
\end{align}
Here, ${\sf n}_\text{r}$ and ${\sf n}_\text{t}$ represent the leading terms of the receive and transmit DOF under isotropic scattering and uncoupled antennas in \eqref{n_rt}, as shown by
\begin{align}  \label{DOF_Landau_omni}
\frac{1}{\pi} \int_0^{\pi/2} \int_0^{2\pi} \cos \theta \, \sin \theta \, d\theta \, d\phi = 2 \int_0^1 y \, dy = 1
\end{align}
using $y=\sin\theta$.
The terms ${\sf n}_\text{r}$ and ${\sf n}_\text{t}$ are multiplied by the respective fractions made available due to the channel selectivity and array orientation at each end of the link, which are tantamount to an angular support reduction and a broadside projection. 
We note that the $\cos(\cdot)$ terms in \eqref{n_r_prime} and \eqref{n_t_prime} result from the change of variables from wavenumber to spherical as per \eqref{wavenumber_spherical} \cite{PoonDoF}. In contrast, the same term in \eqref{variances_channel_composite} arises due to coupling.
For a narrow selectivity and an array oriented such that the scattering is broadside ($\cos \theta_\text{r} \approx 1$),
\begin{align}   \label{n_r_prime_broadside}
{\sf n}_\text{r}^\prime & \approx  {\sf n}_\text{r}  \cdot \frac{2}{\pi} \, \left|{\rm supp}(\Ex\{|\tilde{H}^+_\text{r}|^2\})\right|
\end{align}
at the receiver.
Instead, for an array oriented such that the scattering is at the endfire direction ($\cos \theta_\text{r} \approx \theta_\text{r} - \pi/2$),
\begin{align}   \label{n_r_prime_endfire}
{\sf n}_\text{r}^\prime & \approx  {\sf n}_\text{r}  \cdot \frac{1}{\pi} \!\iint_{{\rm supp}(\Ex\{|\tilde{H}^+_\text{r}|^2\})} \!\!\!  (\pi/2 - \theta_\text{r}) \, \sin \theta_\text{r} \, d\theta_\text{r} \, d\phi_\text{r}.
\end{align}
%
Due to the antenna pattern boundedness, the support is unchanged by coupling: ${\rm supp}(\Ex\{|\tilde{{\sf H}}^+_\text{t}|^2\}) = {\rm supp}(\Ex\{|\tilde{{H}}^+_\text{t}|^2\})$. Transmit coupling is thus immaterial to ${\sf n}^\prime_\text{t}$, which 
hinges solely on the scattering selectivity and array orientation. (As mention at the onset of the section, DOF beyond the uncoupled limit at ${\sf n}_{\text t}$ are excluded from this analysis.) 

Backing off from ${\sf SNR} \to \infty$, the spectral supports become SNR-dependent. By strengthening the eigenvalue polarization (see Fig.~\ref{fig:channel_var_coupled_iso}), coupling can increase the number of spatial dimensions that are usable at high---but finite---SNRs. 
Put differently, coupling cannot enhance the asymptotic slope of the capacity versus log-SNR function, but it can steepen that slope at high---but finite---SNRs. Besides the non-asymptotic slope, coupling is certain to affect the power offset, the zero-order term in the refinement of the expansion in \eqref{C_SNR_high}. The power offset anchors the expansion, and follow-up work is needed to quantify how coupling affects it.
{\color{blue}

\subsection{Impact of Coupling on the Ergodic Capacity}

\begin{figure}
\centering\vspace{-0.0cm}
\includegraphics[width=.999\linewidth]{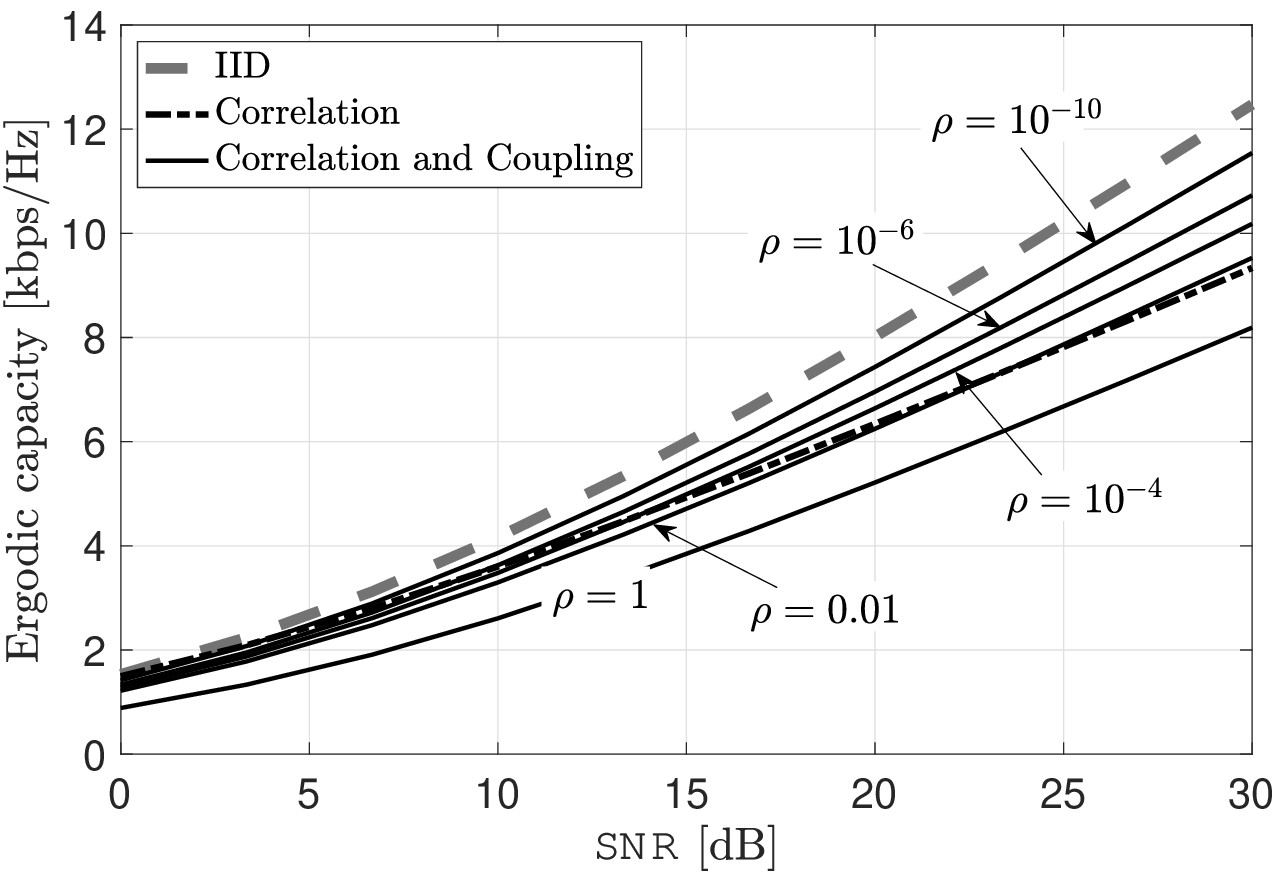} 
\caption{Ergodic capacity of the exact channel versus SNR for various $\rho$ under isotropic scattering at transmitter and IID fading at receiver. UPAs with apertures $15 \lambda$ and antenna spacing $0.4 \lambda$ along each planar dimension. The transmit antennas have an omnidirectional pattern (decorrelating the fading) while the receive antennas are uncoupled.}
\label{fig:capacity_highSNR}
\end{figure}

To gauge the impact of eigenvalue polarization and DOF augmentation on the capacity, the exact $\vect{{\sf H}}$ in \eqref{channel_samples_coupled} is applied. 
Communication occurs between square UPAs of apertures $15 \lambda$ with antennas spaced by $0.4 \lambda$ along each planar dimension. 
IID fading at the receiver emphasizes the interplay between coupling and correlation at the transmitter, yielding, from \eqref{channel_samples_coupled},
 \begin{align} \label{channel_samples_coupled_iidrx}
\vect{{\sf H}}(\rho) = \sqrt{\frac{2}{{\sf R}}} \, \vect{H}_\text{IID} \vect{R}_\text{t}^{1/2} \vect{{\sf C}}_\text{t}^{-1/2}(\rho)
\end{align}
where $\vect{H}_\text{IID} \in \Complex^{N_\text{r} \times N_\text{t}}$ has IID standard complex Gaussian entries. Here, recall, $\vect{R}_\text{t}$ and $\vect{{\sf C}}_\text{t}(\rho)$ denote the transmit correlation and coupling matrices, as defined in \eqref{corr_exact} and  \eqref{coupling_loss}.
Illustrated in Fig.~\ref{fig:capacity_highSNR} is the ergodic capacity of the exact model, 
obtainable from \eqref{capacity} by substituting $\vect{\tilde{{\sf H}}}$ with $\vect{{\sf H}}(\rho)$ and ${\sf n}_\text{min}$ with $N_\text{min} = \min(N_\text{r}, N_\text{t})$, as a function of ${\sf SNR}$ and for varying $\rho$.
The transmit antennas exhibit an omnidirectional pattern, leading to an optimal high-SNR design through fading decorrelation (recall Fig.~\ref{fig:decorrelation_2a}). 
The capacity curves are benchmarked against the capacity of the IID channel 
and of the uncoupled yet correlated channel (i.e., $\vect{{\sf C}}_\text{t}(0) = \vect{I}_{N_\text{t}}$). 
Coupling-induced decorrelation reduces low-SNR capacity while enhancing high-SNR capacity,  
with the transition between these regions marked by the intersection of the capacity curves with the uncoupled curve in Fig.~\ref{fig:capacity_highSNR}, for every $\rho$; lower loss factors $\rho$ require a smaller SNR for this crossing and asymptotically approach the IID capacity limit as $\rho \to 0$.
Additional simulations reveal that coupling effects amplify with smaller antenna spacings.}

\section{Conclusion} \label{sec:conclusion}

An informed holographic MIMO transmitter can optimize its channel capacity across varying SNR levels by accounting for correlation and mutual coupling in the precoder design. The configurability of the antenna patterns would add flexibility, 
allowing the transmitter to reinforce antenna correlations at low SNR 
and 
tone them down at high SNR
so as to optimize the usability of the spatial dimensions provided by the scattering environment and array apertures. In essence, the adjustment of the antenna patterns mimics a reshaping of the environmental scattering. 

Potential directions for future work include the following.
\begin{itemize}
\item
Accounting for mutual coupling at the receiver, due to finite load impedances at those antenna ports.
\item
Incorporating matching networks into the circuit model. 
\item
Evaluating the impact of coupling on the power offset, the zero-order term of the capacity expansion at high SNR. 
\item
Identifying antenna patterns that maximize channel capacity across all SNR levels, generalizing the guidelines obtained for low and high SNR.
\item
Determining the optimal antenna spacing for given losses, patterns, and scattering conditions.
\item
Generalizing the framework to wideband channels \cite{Heath2023}, addressing the superposition of time-harmonic currents and voltages across the communication bandwidth.
\end{itemize}

Also, an inherent assumption throughout the paper has been that all of the antennas within the transmit array
are identical.
By relaxing this condition, the impact of coupling could be extended to the realm of pattern and/or polarization diversity, in which antennas are purposely exposed to distinct portions of the channel spectrum and/or polarization as a mechanism to diminish their correlation \cite{1216759}.



\appendices

\begin{figure*}[t]
\begin{align}  \label{bb0} \tag{128}
z_{\text{t,t}}(\vect{{\sf r}}-\vect{{\sf s}})   
& =  \frac{\kappa Z_0}{8\pi^2} 
\iint_{-\infty}^\infty d\vect{\kappa} \, \frac{1}{\gamma(\vect{\kappa})}   \iiint_{-\infty}^\infty d\vect{{\sf p}} \, {\sf a}_\text{t}(\vect{{\sf p}}-\vect{{\sf r}}) e^{\imagunit \vect{\kappa}^{\Ttran} \vect{p}} \iiint_{-\infty}^\infty d\vect{{\sf q}}  \,  {\sf a}_\text{t}(\vect{{\sf q}}-\vect{{\sf s}}) e^{-\imagunit \vect{\kappa}^{\Ttran} \vect{q}}  e^{\imagunit \gamma |p_z-q_z|}   \\ \label{bb1} \tag{129}
& =  \frac{\kappa Z_0}{8\pi^2} 
\iint_{-\infty}^\infty d\vect{\kappa} \,  \frac{e^{\imagunit \vect{\kappa}^{\Ttran} (\vect{r}-\vect{s})}}{\gamma(\vect{\kappa})} 
  \int_{-\infty}^\infty dp_z \, \overline{{\sf A}_\text{t}(\vect{\kappa},p_z-r_z)}
   \int_{-\infty}^\infty dq_z  \,  {\sf A}_\text{t}(\vect{\kappa},q_z-s_z)  
 e^{\imagunit \gamma |p_z-q_z|} 
 \\ \label{bb3} \tag{130}
 & =  \frac{\kappa Z_0}{8\pi^2} 
\iint_{-\infty}^\infty d\vect{\kappa} \,  \frac{e^{\imagunit \vect{\kappa}^{\Ttran} (\vect{r}-\vect{s})}}{\gamma(\vect{\kappa})} 
  \int_{r-r_0}^{r+r_0} dp_z \, \overline{{\sf A}_\text{t}(\vect{\kappa},p_z-r_z)}
   \int_{s-r_0}^{s+r_0} dq_z  \,  {\sf A}_\text{t}(\vect{\kappa},q_z-s_z)  
 e^{\imagunit \gamma |p_z-q_z|} 
 \end{align} 
 \hrule
\begin{align}   \label{z_impedance} \tag{131}
z_{\text{t,t}}(\vect{{\sf r}}-\vect{{\sf s}})  
 & =  \frac{\kappa Z_0}{2} 
\iint_{-\infty}^\infty \frac{d\vect{\kappa}}{(2\pi)^2} \,  \frac{e^{\imagunit \vect{\kappa}^{\Ttran} (\vect{r}-\vect{s})}}{\gamma(\vect{\kappa})} 
\cdot
\begin{cases} \displaystyle
|{\sf A}_\text{t}(\vect{\kappa},\gamma)|^2  \, e^{\imagunit \gamma (r_z-s_z)}, & r_z-s_z>2 r_0 \\ \displaystyle
|{\sf A}_\text{t}(\vect{\kappa},-\gamma)|^2  \, e^{-\imagunit \gamma (r_z-s_z)}, & r_z-s_z<-2 r_0
\end{cases}
\end{align} 
\hrule
\end{figure*} 

\section*{Appendix}
\subsection{Circuit Power and Electromagnetic Power} \label{app:em_circuit_power}

The instantaneous circuit power expended by the continuous multiport system is given by 
\begin{align}   \label{power_circuit_insta}
{\sf P}_{\text c}(t)  & = \iiint_{-\infty}^\infty  j_{\text t}(t,\vect{{\sf r}}) \,v_{\text t}(t,\vect{{\sf r}}) d\vect{{\sf r}}.
\end{align} 
For a time-harmonic source, expressing the circuit quantities in terms of phasors at frequency $\omega$, we have that
\begin{align}   \label{power_circuit_insta_phasor}
{\sf P}_{\text c}(t)  & = \iiint_{-\infty}^\infty  \Re\left\{j_{\text t}(\vect{{\sf r}}) e^{-\imagunit \omega t}\right\} \, \Re\left\{v_{\text t}(\vect{{\sf r}}) e^{-\imagunit \omega t}\right\} d\vect{{\sf r}}.  
\end{align} 
Time-averaging \eqref{power_circuit_insta_phasor}, the transmit power emerges as
\begin{align}    
{\sf P}_{\text c} &= \lim_{T\to\infty} \frac{1}{T} \int_{-T/2}^{T/2} {\sf P}_{\text c}(t) \, dt \\& \label{time_avg_power_electrical}
 = \frac{\omega}{2\pi} \int_{0}^{2\pi/\omega} {\sf P}_{\text c}(t) \, dt \\ \label{circuit_power_avg}
& = \frac{1}{2} \Re\left\{ \iiint_{-\infty}^\infty  \overline{j_{\text t}(\vect{{\sf r}})} \, v_{\text t}(\vect{{\sf r}}) \, d\vect{{\sf r}}\right\}
\end{align}
after using the identity \cite[Eq.~2.347]{PlaneWaveBook}
\begin{align}
\frac{\omega}{\pi} \int_{0}^{2\pi/\omega}  \Re\left\{a_\omega e^{-\imagunit \omega t}\right\} \Re\left\{b_\omega e^{-\imagunit \omega t}\right\} \, dt = \Re\left\{\overline{a_\omega} \, b_\omega\right\}.
\end{align}
Expanding $v_{\text t}(\vect{{\sf r}})$ in \eqref{circuit_power_avg} according to \eqref{voltage} yields \eqref{circuit_power}.
The instantaneous electromagnetic power exerted by a unipolarized space-time current density $\vect{j}_{\text t}(t,\vect{{\sf r}}) = \hat{\vect{j}} j_{\text t}(t,\vect{{\sf r}})$ on a field distribution $\vect{e}_{\text t}(t,\vect{{\sf r}}) = \hat{\vect{j}} e_{\text t}(t,\vect{{\sf r}})$ is  \cite[Eq.~2.127]{PlaneWaveBook} 
\begin{equation}
{\sf P}_{\text{em}}(t) = \iiint_{-\infty}^\infty  j_{\text t}(t,\vect{{\sf r}}) e_{\text t}(t,\vect{{\sf r}}) d\vect{{\sf r}}
\end{equation}
and, for a time-harmonic source specifically,
\begin{align}    \label{wave_power_punctiform}
{\sf P}_\text{em}  & = \frac{1}{2} \Re\left\{ \iiint_{-\infty}^\infty  \overline{j_{\text t}(\vect{{\sf r}})} \, e_{\text t}(\vect{{\sf r}}) \, d\vect{{\sf r}}\right\}
\end{align}
which is expressed in terms of the corresponding phasors.
The field obeys the scalar Helmholtz equation
\cite{ChewBook,PlaneWaveBook}
\begin{equation} \label{Helmholtz}
\nabla^2 e_{\text t}(\vect{{\sf r}}) + \kappa^2 e_{\text t}(\vect{{\sf r}}) =  \imagunit \kappa Z_0 j_{\text t}(\vect{{\sf r}})
\end{equation}
where ${Z_0 \approx 120 \pi}$ is the wave impedance of free-space while 
$\kappa = 2\pi/\lambda$ given $\lambda$ the wavelength.
From the linearity of \eqref{Helmholtz},
\begin{equation} \label{field_no_coupling}
e_{\text t}(\vect{{\sf r}}) = - \imagunit \kappa Z_0 \iiint_{-\infty}^\infty  j_{\text t}(\vect{{\sf s}}) g(\vect{{\sf r}} - \vect{{\sf s}}) d\vect{{\sf s}} ,
\end{equation}
where $g(\vect{{\sf r}})$ is the Green's function \eqref{Green} solving $\nabla^2 g(\vect{{\sf r}}) + \kappa^2 g(\vect{{\sf r}}) =  - \delta(\vect{{\sf r}})$.
Substituting \eqref{field_no_coupling} into \eqref{wave_power_punctiform} yields  \eqref{wave_power_final}.

\subsection{Transmit Power} \label{app:power}

Rewrite \eqref{circuit_power} as an inner product in $L^2$,
\begin{align} \label{pt_conj}
{\sf P}_\text{t} & =  \Re \langle \mathcal{Z}_{\text{t,t}} j_\text{t}, j_\text{t}\rangle  = \frac{1}{2} \langle \mathcal{Z}_{\text{t,t}} j_\text{t}, j_\text{t}\rangle + \frac{1}{2} \overline{\langle \mathcal{Z}_{\text{t,t}} j_\text{t}, j_\text{t}\rangle}.
\end{align}
Then, using the conjugate symmetry of the inner product with the nonconjugate symmetry and self-adjointness of $\mathcal{Z}_{\text{t,t}}$, 
\begin{align} 
\overline{\langle \mathcal{Z}_{\text{t,t}} j_\text{t}, j_\text{t}\rangle}  =  \langle j_\text{t},  \mathcal{Z}_{\text{t,t}} j_\text{t} \rangle  
& =  \langle  {\mathcal{Z}_{\text{t,t}}}^* j_\text{t}, j_\text{t} \rangle \\ \label{conj}
& =  \langle  \overline{\mathcal{Z}_{\text{t,t}}} j_\text{t}, j_\text{t} \rangle.
\end{align}
Plugging \eqref{conj} into \eqref{pt_conj} yields, by linearity,
\begin{align} \label{Pt_operator}
{\sf P}_\text{t} & = \frac{1}{2} \langle \mathcal{Z}_{\text{t,t}} j_\text{t}, j_\text{t} \rangle + \frac{1}{2} \langle  \overline{\mathcal{Z}_{\text{t,t}}} j_\text{t}, j_\text{t} \rangle 
=  \langle \Re\{\mathcal{Z}_{\text{t,t}}\} j_\text{t}, j_\text{t} \rangle
\end{align}
with $\Re\{\mathcal{Z}_{\text{t,t}}\} = (\mathcal{Z}_{\text{t,t}} + \overline{\mathcal{Z}_{\text{t,t}}})/2$ the operator associated with the real part of the impedance kernel.

\subsection{Transmit Impedance with Physical Antennas} \label{app:impedance_kernel} 

Replacing $g(\vect{\cdot})$ in \eqref{bb} with its spectral representation via Weyl's identity in \eqref{Weyl} yields \eqref{bb0}. Here, the $z$-axis is aligned to the axis connecting the centroids of any two antennas, as shown in Fig.~\ref{fig:impedance_corr_tot}. This transformation is allowed by the rotational invariance of \eqref{Weyl}. 
Taking the 2D Fourier transform on a plane orthogonal to $z$ gives \eqref{bb1}.
Reflecting the space limitation of physical antennas into the integration region leads to \eqref{bb3}. 
Assume $r_z > s_z > 0$. With $p_z = r_z-r_0$ and $q_z=s_z+r_0$, it follows that $|p_z-q_z| = |r_z-s_z-2 r_0|$. Then, $|p_z-q_z|=p_z-q_z$ for $r_z-s_z > 2 r_0$, corresponding to causal antennas; see Fig.~\ref{fig:impedance_corr_tot}. Similarly, assume now $s_z > r_z > 0$. With $p_z = r_z+r_0$ and $q_z=s_z-r_0$, it follows that $|p_z-q_z| = |r_z-s_z+2 r_0|$. Thus, $|p_z-q_z| = q_z-p_z$ for $r_z-s_z <- 2 r_0$, implying anticausal antennas.
In these regimes, \eqref{bb3} simplifies to \eqref{z_impedance} as per translation property of the Fourier transform. 
Generalizing \eqref{z_impedance} to arbitrary rotations leads to the transmit impedance reported in \eqref{impedance_kernel_spectral_noniso}.

\subsection{Normalization of the Coupling Kernel} \label{app:normalization_impedance}

Changing of variables $\vect{{\sf t}} = \vect{{\sf p}}-\vect{{\sf r}}$ and $\vect{{\sf y}} = \vect{{\sf q}}-\vect{{\sf s}}$ in \eqref{bb} while defining $\vect{{\sf v}} = \vect{{\sf r}}-\vect{{\sf s}}$,
\setcounter{equation}{131}
\begin{align}   \label{cc}
z_{\text{t,t}}(\vect{{\sf v}})  & = -\imagunit \kappa Z_0  \iiint_{-\infty}^\infty \!\!\! d\vect{{\sf t}}  \, {\sf a}_\text{t}(\vect{{\sf t}}) \iiint_{-\infty}^\infty \!\!\! d\vect{{\sf y}} \, g(\vect{{\sf v}} + \vect{{\sf t}}-\vect{{\sf y}}) {\sf a}_\text{t}(\vect{{\sf y}}).
\end{align}
Recalling the Green's function spectrum in \cite[Eq.~2.2.23]{ChewBook},
\begin{align} \label{Green_fourier}
G(\vect{{\sf k}})  & = \frac{1}{\|\vect{{\sf k}}\|^2 - \kappa^2} = \frac{1}{(\kappa_z - \gamma)(\kappa_z + \gamma)},
\end{align}
we have that
\begin{align} \label{Green_fourier_shifted}
g(\vect{{\sf v}} + \vect{{\sf t}}-\vect{{\sf y}}) = \iiint_{-\infty}^\infty \frac{d\vect{{\sf k}}}{(2\pi)^3}  \,  G(\vect{{\sf k}}) e^{\imagunit \vect{{\sf k}}^{\Ttran} (\vect{{\sf t}}-\vect{{\sf y}})}   \, e^{\imagunit \vect{{\sf k}}^{\Ttran} \vect{{\sf v}}}.
\end{align}
Substituting \eqref{Green_fourier_shifted} into \eqref{cc} yields
\begin{align} \label{cc1}
z_{\text{t,t}}(\vect{{\sf v}})  & = -\imagunit \kappa Z_0  \iiint_{-\infty}^\infty \frac{d\vect{{\sf k}}}{(2\pi)^3}  \, |{\sf A}_\text{t}(\vect{{\sf k}})|^2 G(\vect{{\sf k}})  \, e^{\imagunit \vect{{\sf k}}^{\Ttran} \vect{{\sf v}}} 
\end{align}
which is due to Hermitian symmetry ${\sf A}_{\text t}(-\vect{{\sf k}}) = \overline{{\sf A}_{\text t}(\vect{{\sf k}})}$, as per ${\sf a}_\text{t}(\vect{{\sf r}})$ being a real-valued function. 
Comparing \eqref{cc1} against \eqref{normalization}, the real part of the transmit impedance reads as
\begin{align}
{\sf c}_\text{t}(\vect{{\sf v}})  & = \Re\left\{- \imagunit \frac{4 \pi}{\kappa}  \!\! \iiint_{-\infty}^\infty  \! \frac{d\vect{{\sf k}}}{(2\pi)^3}  \, |{\sf A}_\text{t}(\vect{{\sf k}})|^2 G(\vect{{\sf k}})  \, e^{\imagunit \vect{{\sf k}}^{\Ttran} \vect{{\sf v}}} \right\}.
\end{align}
Focusing on $\vect{{\sf v}}=\vect{{\sf 0}}$, the transmit self-impedance is given by
\begin{align} 
{\sf c}_\text{t}(\vect{{\sf 0}})  & = \Re\left\{- \frac{\imagunit}{2\pi^2 \kappa}  \iiint_{-\infty}^\infty d\vect{{\sf k}}  \, |{\sf A}_\text{t}(\vect{{\sf k}})|^2 G(\vect{{\sf k}})  \right\} \\ \label{c0}
& = \Re\left\{- \frac{\imagunit}{2\pi^2 \kappa}  \iiint_{-\infty}^\infty d\vect{{\sf k}}  \, \frac{|{\sf A}_\text{t}(\vect{{\sf k}})|^2}{(\kappa_z - \gamma)(\kappa_z + \gamma)} \right\}
\end{align}
in light of \eqref{Green_fourier}.
The integration over $\kappa_z$ reveals a nonzero real part of the impedance. Specifically, for an analytic function $f(z)$ with a simple pole $z_0 \in \Real$, crossing the singularity replaces the integral with its Cauchy principal value, yielding
\begin{align} \label{prova}
\mathrm{PV} \int_{-\infty}^\infty dz  \,  \frac{f(z)}{z - z_0}  =  \imagunit \pi f(z_0)
\end{align}
where the integral exists.
Rewriting the integrand of \eqref{c0} as its partial fraction decomposition 
and integrating over $\kappa_z$,
\begin{align}  \nonumber
& \mathrm{PV}  \int_{-\infty}^\infty d\kappa_z  \, \frac{|{\sf A}_\text{t}(\vect{{\sf k}})|^2}{(\kappa_z - \gamma)(\kappa_z + \gamma)}  \\ \label{PV1}
& \hspace{.5cm} =   \mathrm{PV}
\int_{-\infty}^\infty d\kappa_z  \, \left( \frac{|{\sf A}_\text{t}(\vect{{\sf k}})|^2}{2 \gamma (\kappa_z - \gamma)} + \frac{|{\sf A}_\text{t}(\vect{{\sf k}})|^2}{2 \gamma (\kappa_z + \gamma)} \right) \\ \label{PV}
& \hspace{.5cm} = \imagunit \pi \left(\frac{|{\sf A}_\text{t}^+(\vect{\kappa})|^2}{2\gamma}  + \frac{|{\sf A}_\text{t}^-(\vect{\kappa})|^2}{2\gamma} \right)
\end{align}
with the spectra ${\sf A}_\text{t}^\pm(\vect{\cdot})$ as defined in \eqref{impedance_kernel_spectral_noniso}.
Note that the above integration requires the poles $\kappa_z = \pm \gamma$ to lie on the real axis, a condition satisfied for $\|\vect{k}\|\le\kappa$, where $\gamma \in \Real$ as per \eqref{gamma}.
Finally, substituting \eqref{PV} into \eqref{c0},
\begin{align} \label{c2}
{\sf c}_\text{t}(\vect{{\sf 0}})  
& = \frac{1}{4 \pi \kappa}  \iint_{\|\vect{\kappa}\|\le\kappa} d\vect{\kappa}  \, \frac{|{\sf A}_\text{t}^+(\vect{\kappa})|^2  + |{\sf A}_\text{t}^-(\vect{\kappa})|^2}{\gamma(\vect{\kappa})},
\end{align}
which leads to the normalization condition in \eqref{norm_A_spectrum}.

\subsection{Existence of Deconvolution} \label{app:invertible_Chalf}
 
 The invertibility of $\mathcal{C}^{1/2}$ is proven next by contradiction. 
 Assume that ${\sf P}_\text{t} > 0$, thereby making $\mathcal{C}$ positive-definite as per \eqref{psd_op}, and that $|{\sf A}_{\text{t}}(\vect{{\sf k}})|$ is only nonnegative, meaning this function could be zero in a non-empty region $K\subset \Real^3$. Then, a current spectrum $J_{\text t}(\vect{{\sf k}})$ could be chosen to be zero everywhere except within $K$, whereby the associated power density ${\sf S}_{\text t}(\vect{{\sf k}})$ in \eqref{psd_Dirac_LSI_noniso} would be zero $\forall \vect{{\sf k}}$, implying ${\sf P}_\text{t} = 0$ as per \eqref{power_psd}. However, this is not allowed given that ${\sf P}_\text{t} > 0$ by hypothesis.

\subsection{Maximum Eigenvalue} \label{app:max_eigenvalue}

For a single channel realization, the maximum eigenvalue of $\tilde{\vect{{\sf H}}}$ in \eqref{equiv_channel_coupling} equals its spectral norm. Repeatedly applying the sub-multiplicativity property of a matrix norm,
\begin{align}
\lambda_\text{max}(\tilde{\vect{{\sf H}}} \tilde{\vect{{\sf H}}}^{\Htran})
& = \|\tilde{\vect{{\sf H}}}\|_2^2 \\
& \approx \| \vect{\Lambda}_{\text{r}}^{1/2} \vect{W} \vect{{\sf \Lambda}}_{\text{t}}^{1/2} \|_2^2 \\& 
\le \|\vect{\Lambda}_{\text{r}}^{1/2} \|_2^2  \|\vect{W} \|_2^2 \| \vect{{\sf \Lambda}}_{\text{t}}^{1/2} \|_2^2   \\&
= \| \vect{\Lambda}_{\text{r}} \|_2  \|\vect{W} \|_2^2 \| \vect{{\sf \Lambda}}_{\text{t}} \|_2   \\& \label{lambda_HC}
=  \lambda_\text{max}(\vect{\Lambda}_{\text{r}}) \lambda_\text{max}(\vect{W} \vect{W}^{\Htran}) \lambda_\text{max}(\vect{{\sf \Lambda}}_{\text{t}}),
\end{align}
where $\lambda_\text{max}(\vect{\Lambda}_{\text{r}})$ and $\lambda_\text{max}(\vect{{\sf \Lambda}}_{\text{t}})$ denote the maximum eigenvalues of the receive and transmit correlations, given by
\begin{align}  \label{eig_rx_max}
\lambda_\text{max}(\vect{\Lambda}_{\text{r}}) & = N_{\text{r}} \, \max_{\vect{i} \in \Lambda_\text{r}} \sigma^2_{\vect{i}}(\tilde{H}_\text{r}^+)  \\ \label{eig_tx_max}
\lambda_\text{max}(\vect{{\sf \Lambda}}_{\text{t}}) & =  N_{\text{t}} \, \max_{\vect{j} \in \Lambda_\text{t}} \sigma^2_{\vect{j}}(\tilde{{\sf H}}_\text{t}^+).
\end{align}
The inequality \eqref{bound_eig_HC_discrete} is derived by averaging \eqref{lambda_HC} over all channel realizations while substituting \eqref{eig_rx_max} and \eqref{eig_tx_max}.

\subsection{Spatial DOF} \label{app:DOF}

The asymptotic slope of the spectral efficiency is \cite{LozanoCorrelation}
\begin{align}
{\sf DOF} = \lim_{{\sf SNR}\to \infty} \frac{{\sf SNR}}{\log_2 e} \, \frac{d{\sf C}({\sf SNR})}{d{\sf SNR}}.
\end{align}
Substituting ${\sf C}({\sf SNR})$ from \eqref{capacity} yields \cite[Appendix~G]{LozanoCorrelation}
\begin{equation}
{\sf DOF} = \Ex \{ \rank(\tilde{\vect{{\sf H}}} \vect{{\sf P}}^\star \tilde{\vect{{\sf H}}}^{\Htran}) \}
\end{equation}
and, as at high SNR
waterfilling allocates power onto every direction associated with a nonzero eigenvalue of $\tilde{\vect{{\sf H}}} \tilde{\vect{{\sf H}}}^{\Htran}$  
\begin{equation} \label{S_inf_UIU}
{\sf DOF} = \Ex \! \left \{ \rank(\tilde{\vect{{\sf H}}}^\prime (\tilde{\vect{{\sf H}}}^\prime)^{\Htran}) \right \},
\end{equation}
where $\tilde{\vect{{\sf H}}}^\prime$ is the ${\sf DOF}_\text{r}^\prime \times {\sf DOF}_\text{t}^\prime$ submatrix (${\sf DOF}_\text{r}^\prime \le {\sf DOF}_\text{r}$ and ${\sf DOF}_\text{t}^\prime \le {\sf DOF}_\text{t}$) obtained by removing the rows and columns tied to zero eigenvalues of $\tilde{\vect{{\sf H}}}$.
As the entries of $\tilde{\vect{{\sf H}}}^\prime$ are independent, its rows and columns are linearly independent with probability~$1$ save for those identically zero, resulting in \eqref{S_inf_UIU_rank}.
For the separable model in \eqref{equiv_channel_coupling}, ${\sf DOF}_\text{r}^\prime$ and ${\sf DOF}_\text{t}^\prime$ 
correspond to the cardinalities of the 2D lattices 
\begin{align} \nonumber
\Lambda_\text{r}^\prime & = \Big\{\vect{i} \in \Integer^2 :  \{\|\vect{k}\| \le \kappa\} \bigcap \left\{\|\vect{D}_\text{r} \vect{k} - \kappa \vect{i}\|_\infty\le \tfrac{\kappa}{2} \right\} \\& \hspace{3cm} \label{lattice_rx_prime}
\bigcap {\rm supp}(\Ex\{|\tilde{H}^+_\text{r}|^2\}) \neq \emptyset \Big\} \\ \nonumber
\Lambda_\text{t}^\prime & = \Big\{\vect{j} \in \Integer^2 :  \{\|\vect{\kappa}\| \le \kappa\} \bigcap \left\{\|\vect{D}_\text{t} \vect{\kappa} - \kappa \vect{j}\|_\infty\le \tfrac{\kappa}{2} \right\} \\& \hspace{3cm} \label{lattice_tx_prime}
\bigcap {\rm supp}(\Ex\{|\tilde{{\sf H}}^+_\text{t}|^2\}) \neq \emptyset \Big\},
\end{align}
where the wavenumber disk is defined by stationarity, ${\rm supp}(\cdot)$ accounts for the scattering selectivity, and $\{\|\bm{\cdot}\|_\infty \le \tfrac{\kappa}{2}\}$ 
models the impact of array apertures \cite{PizzoTWC21}.
Normalization by $\kappa$ keeps the DOF unchanged in \eqref{lattice_rx_prime} and \eqref{lattice_tx_prime} while emphasizing their dependance on $\lambda$. With this applied normalization, at the receiver,
as $L_{\text{r},x}/\lambda$ increases, each set contracts, and \eqref{lattice_rx_prime} approximates the product of the packing density and wavenumber support:
\begin{align} \label{n_r_prime_wave_0}
{\sf n}_\text{r}^\prime & =  \left\lceil \det(\vect{D}_\text{r}) \, \iint_{\{\|\vect{k}\| \le 1\} \, \bigcap \, {\rm supp}(\Ex\{|\tilde{H}^+_\text{r}|^2\})} \, d\vect{k} \right\rceil \\ \label{n_r_prime_wave}
& =  \left\lceil {\sf n}_\text{r} \cdot \frac{1}{\pi} \iint_{\{\|\vect{k}\| \le 1\} \, \bigcap \, {\rm supp}(\Ex\{|\tilde{H}^+_\text{r}|^2\})} \, d\vect{k} \right\rceil,
\end{align}
subsuming ${\sf n}_\text{r}$ in \eqref{n_rt} under isotropic scattering. 
The transmit formula is derived analogously to \eqref{n_r_prime_wave}, but with a coupling-inclusive power spectrum as
\begin{align} \label{n_t_prime_wave} 
{\sf n}_\text{t}^\prime & =  \left\lceil {\sf n}_\text{t} \cdot \frac{1}{\pi} \iint_{\{\|\vect{\kappa}\| \le 1\} \, \bigcap \, {\rm supp}(\Ex\{|\tilde{{\sf H}}^+_\text{t}|^2\})} \, d\vect{\kappa} \right\rceil.
\end{align}
From \eqref{n_r_prime_wave} and \eqref{n_t_prime_wave}, shifting from wavenumber representation (with axes rescaled by $\kappa$) to spherical as per \eqref{wavenumber_spherical} yields \eqref{n_r_prime} and \eqref{n_t_prime}.

\bibliographystyle{IEEEtran}
\bibliography{IEEEabrv,refs}

\begin{thebibliography}{10}
\providecommand{\url}[1]{#1}
\csname url@samestyle\endcsname
\providecommand{\newblock}{\relax}
\providecommand{\bibinfo}[2]{#2}
\providecommand{\BIBentrySTDinterwordspacing}{\spaceskip=0pt\relax}
\providecommand{\BIBentryALTinterwordstretchfactor}{4}
\providecommand{\BIBentryALTinterwordspacing}{\spaceskip=\fontdimen2\font plus
\BIBentryALTinterwordstretchfactor\fontdimen3\font minus
  \fontdimen4\font\relax}
\providecommand{\BIBforeignlanguage}[2]{{%
\expandafter\ifx\csname l@#1\endcsname\relax
\typeout{** WARNING: IEEEtran.bst: No hyphenation pattern has been}%
\typeout{** loaded for the language `#1'. Using the pattern for}%
\typeout{** the default language instead.}%
\else
\language=\csname l@#1\endcsname
\fi
#2}}
\providecommand{\BIBdecl}{\relax}
\BIBdecl

\bibitem{10144733}
E.~Björnson, Y.~C. Eldar, E.~G. Larsson, A.~Lozano, and H.~V. Poor,
  ``Twenty-five years of signal processing advances for multiantenna
  communications: {From} theory to mainstream technology,'' \emph{IEEE Signal
  Proc. Mag.}, vol.~40, no.~4, pp. 107--117, 2023.

\bibitem{BJORNSON20193}
E.~Bj{\"o}rnson, L.~Sanguinetti, H.~Wymeersch, J.~Hoydis, and T.~L. Marzetta,
  ``Massive {MIMO} is a reality---{W}hat is next?: Five promising research
  directions for antenna arrays,'' \emph{Dig. Signal Process.}, vol.~94, pp.
  3--20, 2019.

\bibitem{PizzoTSP21}
A.~Pizzo, A.~d.~J. Torres, L.~Sanguinetti, and T.~L. Marzetta, ``Nyquist
  sampling and degrees of freedom of electromagnetic fields,'' \emph{IEEE
  Trans. Signal Process.}, vol.~70, pp. 3935--3947, 2022.

\bibitem{Prather2016}
D.~W. Prather, ``Toward holographic {RF} systems for wireless communications
  and networks,'' \emph{IEEE ComSoc Tech. News}, 2016.

\bibitem{JensenReview}
M.~Jensen and J.~Wallace, ``A review of antennas and propagation for {MIMO}
  wireless communications,'' \emph{IEEE Trans. Antennas Propag.}, vol.~52,
  no.~11, pp. 2810--2824, 2004.

\bibitem{PizzoIT21}
A.~Pizzo, L.~Sanguinetti, and T.~L. Marzetta, ``Spatial characterization of
  electromagnetic random channels,'' \emph{IEEE Open J. Commun. Soc.}, vol.~3,
  pp. 847--866, 2022.

\bibitem{PizzoJSAC20}
A.~{Pizzo}, T.~L. {Marzetta}, and L.~{Sanguinetti}, ``Spatially-stationary
  model for holographic {MIMO} small-scale fading,'' \emph{IEEE J. Sel. Areas
  Commun.}, vol.~38, no.~9, pp. 1964--1979, 2020.

\bibitem{PizzoTWC22}
A.~Pizzo, A.~Lozano, S.~Rangan, and T.~L. Marzetta, ``Wide-aperture {LOS}
  {MIMO} via reflections off a smooth surface,'' \emph{IEEE Trans. Wireless
  Commun.}, 2022.

\bibitem{Prather2013}
G.~J. Schneider, J.~A. Murakowski, C.~A. Schuetz, S.~Shi, and D.~W. Prather,
  ``Radiofrequency signal-generation system with over seven octaves of
  continuous tuning,'' \emph{Nature Photonics}, vol.~7, pp. 118--122, 2013.

\bibitem{Prather2017}
D.~W. Prather~{\textit{et al.}}, ``Optically upconverted, spatially coherent
  phased-array-antenna feed networks for beam-space {MIMO} in {5G} cellular
  communications,'' \emph{IEEE Trans. Antennas Propag.}, vol.~65, no.~12, pp.
  6432--6443, 2017.

\bibitem{Insang2019}
I.~Yoo, M.~F. Imani, T.~Sleasman, H.~D. Pfister, and D.~R. Smith, ``Enhancing
  capacity of spatial multiplexing systems using reconfigurable cavity-backed
  metasurface antennas in clustered {MIMO} channels,'' \emph{IEEE Trans.
  Commun.}, vol.~67, no.~2, pp. 1070--1084, 2019.

\bibitem{Steyskal1990}
H.~Steyskal and J.~Herd, ``Mutual coupling compensation in small array
  antennas,'' \emph{IEEE Trans. Antennas Propag.}, vol.~38, no.~12, pp.
  1971--1975, 1990.

\bibitem{Birtcher2006}
Z.~Huang, C.~A. Balanis, and C.~R. Birtcher, ``Mutual coupling compensation in
  {UCA}s: Simulations and experiment,'' \emph{IEEE Trans. Antennas Propag.},
  vol.~54, no.~11, pp. 3082--3086, 2006.

\bibitem{Wallace2005}
M.~Morris, M.~Jensen, and J.~Wallace, ``Superdirectivity in {MIMO} systems,''
  \emph{IEEE Trans. Antennas Propag.}, vol.~53, no.~9, pp. 2850--2857, 2005.

\bibitem{Marzetta2019}
T.~L. Marzetta, ``Super-directive antenna arrays: Fundamentals and new
  perspectives,'' in \emph{53rd Asilomar Conference on Signals, Systems, and
  Computers}, 2019, pp. 1--4.

\bibitem{Matthaiou2023}
K.~Dovelos, S.~D. Assimonis, H.~Q. Ngo, and M.~Matthaiou, ``Superdirective
  antenna pairs for energy-efficient terahertz massive {MIMO},'' \emph{IEEE
  Trans. Commun.}, vol.~71, no.~12, pp. 7316--7332, 2023.

\bibitem{Wallace2004}
J.~Wallace and M.~Jensen, ``Mutual coupling in {MIMO} wireless systems: A
  rigorous network theory analysis,'' \emph{IEEE Trans. Wireless Commun.},
  vol.~3, no.~4, pp. 1317--1325, 2004.

\bibitem{Clerckx2007}
B.~Clerckx, C.~Craeye, D.~Vanhoenacker-Janvier, and C.~Oestges, ``Impact of
  antenna coupling on 2 $\times$ 2 {MIMO} communications,'' \emph{IEEE Trans.
  Vehicular Tech.}, vol.~56, no.~3, pp. 1009--1018, 2007.

\bibitem{Nossek2010}
M.~T. Ivrlač and J.~A. Nossek, ``Toward a circuit theory of communication,''
  \emph{IEEE Trans. Circuits and Systems}, vol.~57, no.~7, pp. 1663--1683,
  2010.

\bibitem{Masouros2013}
C.~Masouros, M.~Sellathurai, and T.~Ratnarajah, ``Large-scale {MIMO}
  transmitters in fixed physical spaces: The effect of transmit correlation and
  mutual coupling,'' \emph{IEEE Trans. Commun.}, vol.~61, no.~7, pp.
  2794--2804, 2013.

\bibitem{Heath2023}
M.~Akrout, V.~Shyianov, F.~Bellili, A.~Mezghani, and R.~W. Heath,
  ``Super-wideband massive {MIMO},'' \emph{IEEE J. Sel. Areas Commun.},
  vol.~41, no.~8, pp. 2414--2430, 2023.

\bibitem{Janaswamy2002}
R.~{Janaswamy}, ``Effect of element mutual coupling on the capacity of fixed
  length linear arrays,'' \emph{IEEE Antennas Wireless Propag. Lett.}, vol.~1,
  pp. 157--160, 2002.

\bibitem{Branislav2024}
P.~Wang, M.~N. Khormuji, and B.~M. Popovic, ``Beamforming performances of
  holographic surfaces,'' \emph{IEEE Trans. Wireless Commun.}, vol.~23, no.~6,
  pp. 5816--5831, 2024.

\bibitem{Sanguinetti2024}
A.~A. D’Amico and L.~Sanguinetti, ``Holographic {MIMO} communications: What
  is the benefit of closely spaced antennas?'' \emph{IEEE Trans. Wireless
  Commun.}, pp. 1--1, 2024.

\bibitem{Sha2023}
S.~S.~A. Yuan, X.~Chen, C.~Huang, and W.~E.~I. Sha, ``Effects of mutual
  coupling on degree of freedom and antenna efficiency in holographic {MIMO}
  communications,'' \emph{IEEE Open J. Antennas Propag.}, vol.~4, pp. 237--244,
  2023.

\bibitem{Tengjiao2022}
W.~Tengjiao~{\textit{et al.}}, ``Electromagnetic-compliant channel modeling and
  performance evaluation for holographic {MIMO},'' in \emph{2022 IEEE Globecom
  Workshops (GC Wkshps)}, 2022, pp. 747--752.

\bibitem{Tengjiao2023}
------, ``Channel measurement for holographic {MIMO}: Benefits and challenges
  of spatial oversampling,'' in \emph{2023 IEEE Int. Conf. Commun. (ICC)},
  2023, pp. 5036--5041.

\bibitem{Yongxi2024}
Y.~Liu, M.~Zhang, T.~Wang, A.~Zhang, and M.~Debbah, ``Densifying {MIMO}:
  Channel modeling, physical constraints, and performance evaluation for
  holographic communications,'' \emph{IEEE J. Sel. Areas Commun.}, vol.~42,
  no.~6, pp. 1504--1518, 2024.

\bibitem{MarzettaIT}
T.~L. Marzetta, E.~G. Larsson, and T.~B. Hansen, ``{Massive MIMO and Beyond},''
  in \emph{Inf. Theoretic Perspectives on 5G Systems and Beyond}, S.~S.
  I.~Maric, O.~Simeone, Ed.\hskip 1em plus 0.5em minus 0.4em\relax Cambridge
  Univ. Press, 2020.

\bibitem{ChewBook}
W.~C. Chew, \emph{Waves and Fields in Inhomogenous Media}.\hskip 1em plus 0.5em
  minus 0.4em\relax Wiley-IEEE Press, 1995.

\bibitem{teal2002spatial}
P.~D. Teal, T.~D. Abhayapala, and R.~A. Kennedy, ``Spatial correlation for
  general distributions of scatterers,'' \emph{IEEE Signal Proc. Letters},
  vol.~9, no.~10, pp. 305--308, 2002.

\bibitem{PlaneWaveBook}
T.~B. Hansen and A.~D. Yaghjian, \emph{Plane-Wave Theory of Time-Domain
  Fields}.\hskip 1em plus 0.5em minus 0.4em\relax New York: Wiley-IEEE Press,
  1999.

\bibitem{Unser1994}
M.~Unser and A.~Aldroubi, ``A general sampling theory for nonideal acquisition
  devices,'' \emph{IEEE Trans. Signal Process.}, vol.~42, no.~11, pp.
  2915--2925, 1994.

\bibitem{PizzoTWC21}
A.~Pizzo, L.~Sanguinetti, and T.~Marzetta, ``Fourier plane-wave series
  expansion for holographic {MIMO} communications,'' \emph{IEEE Trans. Wireless
  Commun.}, pp. 1--1, 2022.

\bibitem{LozanoCorrelation}
A.~M. Tulino, A.~Lozano, and S.~Verdu, ``Impact of antenna correlation on the
  capacity of multiantenna channels,'' \emph{IEEE Trans. Inf. Theory}, vol.~51,
  no.~7, pp. 2491--2509, 2005.

\bibitem{Sayeed2002}
A.~M. {Sayeed}, ``Deconstructing multiantenna fading channels,'' \emph{IEEE
  Trans. Signal Process.}, vol.~50, no.~10, 2002.

\bibitem{heath_lozano_2018}
R.~W. Heath~Jr. and A.~Lozano, \emph{Foundations of {MIMO}
  Communication}.\hskip 1em plus 0.5em minus 0.4em\relax Cambridge University
  Press, 2018.

\bibitem{Franceschetti}
M.~Franceschetti, ``On {L}andau's eigenvalue theorem and information
  cut-sets,'' \emph{IEEE Trans. Inf. Theory}, vol.~61, no.~9, 2015.

\bibitem{chizhik2000effect}
D.~Chizhik, F.~Rashid-Farrokhi, J.~Ling, and A.~Lozano, ``Effect of antenna
  separation on the capacity of {BLAST} in correlated channels,'' \emph{IEEE
  Commun. Letters}, vol.~4, no.~11, pp. 337--339, 2000.

\bibitem{PizzoWCL22}
A.~Pizzo and A.~Lozano, ``On {L}andau's eigenvalue theorem for line-of-sight
  {MIMO} channels,'' \emph{IEEE Wireless Commun. Lett.}, vol.~11, no.~12, pp.
  2565--2569, 2022.

\bibitem{HeedongIRS}
H.~Do, N.~Lee, and A.~Lozano, ``Line-of-sight {MIMO} via intelligent reflecting
  surface,'' \emph{IEEE Trans. Wireless Commun.}, vol.~22, no.~6, pp.
  4215--4231, 2023.

\bibitem{PoonDoF}
A.~S.~Y. Poon, R.~W. Brodersen, and D.~N.~C. Tse, ``Degrees of freedom in
  multiple-antenna channels: a signal space approach,'' \emph{IEEE Trans. Inf.
  Theory}, vol.~51, no.~2, pp. 523--536, Feb 2005.

\bibitem{1216759}
A.~Tulino, S.~Verdu, and A.~Lozano, ``Capacity of antenna arrays with space,
  polarization and pattern diversity,'' in \emph{IEEE Information Theory
  Workshop (ITW)}, 2003, pp. 324--327.

\end{thebibliography}

\end{document}